\documentclass[letterpaper,twocolumn,10pt]{article}
\usepackage{usenix-2020-09}

\usepackage{amsmath,amssymb,amsfonts}
\DeclareMathAlphabet{\mathcal}{OMS}{cmsy}{m}{n} %
\usepackage[shortlabels]{enumitem}
\PassOptionsToPackage{hyphens}{url}\usepackage{hyperref}
\usepackage[normalem]{ulem}
\usepackage{mathtools}
\usepackage{subcaption}
\usepackage{booktabs} %

\usepackage{array}   %
\newcolumntype{L}{>{$}l<{$}} %

\usepackage{bbm}

\usepackage{adjustbox} %
\usepackage[labelfont=bf]{caption} %

\usepackage[ruled, lined, noend, linesnumbered]{algorithm2e}
\SetKwComment{Comment}{$\triangleright$\ }{}

\DontPrintSemicolon

\usepackage{tikz-cd}
\usepackage{tikz}
\usetikzlibrary{calc} 
\usetikzlibrary{decorations.text}
\usetikzlibrary{decorations.pathreplacing}
\usetikzlibrary{arrows}

\usepackage{custom_macros} %

\begin{document}

\date{}

\title{\Large \bf Pool Inference Attacks on Local Differential Privacy: Quantifying the Privacy Guarantees of Apple's Count Mean Sketch in Practice}

 \author{
 {\rm Andrea Gadotti}\\
 Imperial College London
 \and
 {\rm Florimond Houssiau}\\
 \ \ Alan Turing Institute\ \ 
 \and
 {\rm Meenatchi Sundaram Muthu Selva Annamalai}\\
 Imperial College London
 \and
 {\rm Yves-Alexandre de Montjoye\begin{NoHyper}\thanks{Email: \texttt{deMontjoye@imperial.ac.uk}; Corresponding author.}\end{NoHyper}}\\
 Imperial College London
 } %

\maketitle

\begin{abstract}
\renewcommand*{\thefootnote}{\fnsymbol{footnote}}
\setcounter{footnote}{6}
\footnote{This is the extended version of the corresponding peer-reviewed paper that was published at USENIX Security 2022. Please cite this work as: Gadotti, A., Houssiau, F., Annamalai, M.S.M.S., Montjoye, Y.-A. de, 2022. \emph{Pool Inference Attacks on Local Differential Privacy: Quantifying the Privacy Guarantees of Apple’s Count Mean Sketch in Practice.} Presented at the 31st USENIX Security Symposium (USENIX Security 22), pp. 501–518 [\href{https://www.usenix.org/conference/usenixsecurity22/presentation/gadotti}{link}].}
\renewcommand*{\thefootnote}{\arabic{footnote}}
\setcounter{footnote}{0}
Behavioral data generated by users' devices, ranging from emoji use to pages visited, are collected at scale to improve apps and services. These data, however, contain fine-grained records and can reveal sensitive information about individual users. 
Local differential privacy has been used by companies as a solution to collect data from users while preserving privacy. We here first introduce pool inference attacks, where an adversary has access to a user's obfuscated data, defines pools of objects, and exploits the user's polarized behavior in multiple data collections to infer the user's preferred pool. Second, we instantiate this attack against Count Mean Sketch, a local differential privacy mechanism proposed by Apple and deployed in iOS and Mac OS devices, using a Bayesian model. Using Apple's parameters for the privacy loss $\eps$, we then consider two specific attacks: one in the emojis setting --- where an adversary aims at inferring a user's preferred skin tone for emojis --- and one against visited websites --- where an adversary wants to learn the political orientation of a user from the news websites they visit. In both cases, we show the attack to be much more effective than a random guess when the adversary collects enough data. We find that users with high polarization and relevant interest are significantly more vulnerable, and we show that our attack is well-calibrated, allowing the adversary to target such vulnerable users. We finally validate our results for the emojis setting using user data from Twitter. Taken together, our results show that pool inference attacks are a concern for data protected by local differential privacy mechanisms with a large $\eps$, emphasizing the need for additional technical safeguards and the need for more research on how to apply local differential privacy for multiple collections.
\end{abstract}

\section{Introduction}

User's behavioral data, ranging from words typed to processes running on the phone, are collected by operating systems, apps, and services. This data allows companies to better understand user behavior, detect issues, and ultimately improve services. For instance, iOS and Mac OS devices keep track of websites that the user visits using the Safari browser, together with the user's preferences on videos that play automatically when the page is loaded~\cite{appledifferentialprivacyteamLearningPrivacyScale2017}. Aggregated over millions of users, this data allows Apple to learn on which websites the users generally want videos to play automatically and to set default auto-play policies in Safari~\cite{appledifferentialprivacyteamLearningPrivacyScale2017}.

\begin{figure}[tbp]
\centerline{\includegraphics[width=.8\linewidth]{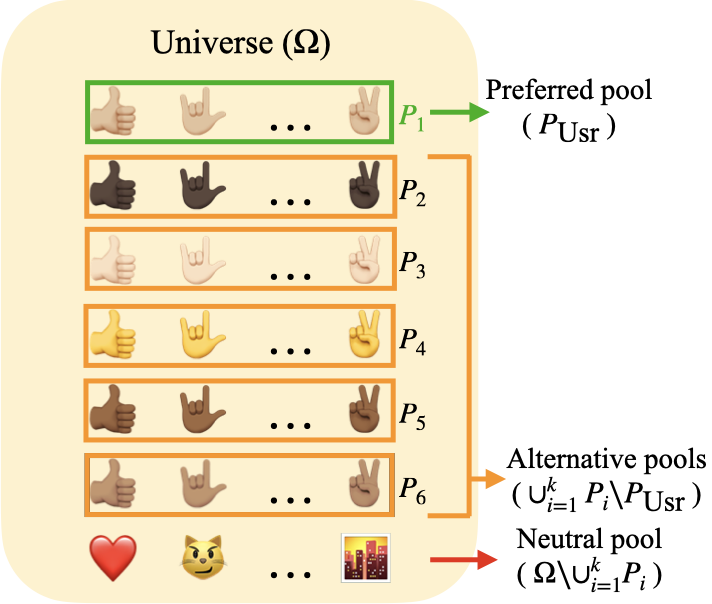}}
\caption{Example of pools defined on a universe $\Omega$ consisting of emojis, when the adversary is interested in determining the skin tone that is most often selected by the user. In this case, $\U$'s preferred pool is the one containing medium-light skin tone emojis.}
\label{fig:skin_tones}
\end{figure}

Local differential privacy, a variation of differential privacy, is among the main solutions for such data collection. Mechanisms satisfying local differential privacy avoid users having to trust anyone, including the data curator. The mechanism takes as input the original data recorded on the user device (original objects) and shares with the curator a randomized version (obfuscated objects) which should not reveal (almost) anything about the original information~\cite{dworkAlgorithmicFoundationsDifferential2013, josephLocalDifferentialPrivacy2018, xiongComprehensiveSurveyLocal2020}. A large literature exists on mechanisms satisfying local differential privacy~\cite{xiongComprehensiveSurveyLocal2020} and some mechanisms have been deployed at scale by Google~\cite{erlingssonRAPPORRandomizedAggregatable2014}, Microsoft~\cite{dingCollectingTelemetryData2017}, and Apple~\cite{appledifferentialprivacyteamLearningPrivacyScale2017}.

One of these mechanisms is Count Mean Sketch (CMS). CMS is used on iOS and Mac OS devices to report both emojis used and websites visited to Apple. Featured in Apple's keynote, local differential privacy allows the company to ``help discover the usage patterns of a large number of users without compromising individual privacy''~\cite{greenbergAppleDifferentialPrivacy}. This implementation, and in particular Apple's choice of the parameter $\eps$, has come under criticism from privacy researchers. It is generally believed that $\eps$ --- which controls the \emph{privacy loss} incurred by the user --- should typically not exceed $\ln(3)$ ($\approx 1.10$)~\cite{dworkDifferentialPrivacyStatistics2010}. Soon after the technology was deployed, it was found that Apple's implementation uses $\eps=4$ when collecting emoji usage data and $\eps=8$ when collecting web domain data~\cite{tangPrivacyLossApple2017}.

Apple's choice to only consider the privacy loss per submission --- once a day for both the web domain and emoji data --- instead of a total privacy loss $\eps_\text{tot}$ (after which objects would no longer be collected from the user~\cite{appleDifferentialPrivacyOverview}) has similarly raised concerns on theoretical ground. While Apple states that they remove user identifiers and IP addresses after the obfuscated objects are received by their server~\cite{appledifferentialprivacyteamLearningPrivacyScale2017}, this is a measure that relies on trust and hence conflicts with local differential privacy's purpose of protecting against an untrusted curator\footnote{If one assumes that Apple removes any identifier --- so that objects from the same user are not linked together and cannot be linked back to individual users ---, then local differential privacy would be mostly unnecessary in the first place. Collecting the original non-obfuscated objects and removing any identifier would already preserve privacy in most settings.}. It is indeed well-known that the mathematical guarantees offered by local differential privacy degrade as multiple objects are collected from the same user, something that can be quantified with an upper bound using the Composition Theorem~\cite{dworkAlgorithmicFoundationsDifferential2013} ($\eps_\text{tot} \leq \eps_1 + \ldots + \eps_n$). Regardless of how revealing the user's original data may be, a low $\eps_\text{tot}$ would guarantee that the obfuscated data will never leak much information. However, $\eps_\text{tot}$ is a worst-case theoretical measure: it is unclear the extent to which collecting multiple objects and using a large $\eps$ for each object open the door to attacks in practice.

\myparagraph{Pool inference attack.}
In this paper we propose the first --- to the best of our knowledge --- quantification of the practical privacy guarantees provided by a deployed local differential privacy mechanism. We design a novel attack against CMS --- which we call \emph{pool inference attack} --- that works as follows: first, the adversary receives a sequence of obfuscated objects from a user; second, the adversary defines pools of interest for the attack (i.e.\ disjoint groups of objects); third, the adversary runs the attack to determine the user's preferred pool --- i.e.\ the pool whose objects are most likely to be selected by the user --- along with a confidence score for the inference. In our first use case, the adversary defines the pools to be groups of emojis divided by skin tone (see Figure~\ref{fig:skin_tones}), the goal of the attack being then to infer which is the emoji skin tone used most frequently by the user.

\myparagraph{Contributions.}
We make the following contributions: \emph{(i)}~We propose pool inference attacks, a new class of attacks aiming at quantifying the sensitive information leaked by local differential privacy mechanisms in practice. We formalize the attack model as a game which can be applied to any mechanism that obfuscates objects independently. \emph{(ii)}~We propose a general Bayesian model for pool inference attacks that can be adapted to most local differential privacy mechanisms. The attack uses a hierarchical probability model that simultaneously encodes properties of the user's behavior, the obfuscation of the mechanism, and auxiliary information that may be available to the adversary. \emph{(iii)}~We instantiate the attack against synthetic users in two practical settings where the adversary's goal is to infer user preferences (1) for emoji skin tone or (2) political news website. We study the impact that properties of user behavior --- such as polarization --- have on the attack's effectiveness, and show that our attack can estimate the probability that its output is correct. We also show that, in some cases, CMS provides little protection compared to a scenario where the user simply submits the true object without any local differential privacy. \emph{(iv)} We simulate the attack in the emojis setting using data from Twitter, and find it to be very effective on users who frequently select emojis supporting skin tones. \emph{(v)}~We discuss potential solutions and mitigation strategies that may prevent our attack or make it less effective.

\section{Background} \label{sec:background}

We now define local differential privacy and the CMS algorithm, introducing the notation that will be used in the paper. 

\myparagraph{Local differential privacy~\cite{kasiviswanathanWhatCanWe2008}.} 
A local differential privacy mechanism is a randomized algorithm that takes as input an \emph{original object} from a set $\Omega$ and returns an \emph{obfuscated object} from a set $\Y$. For example, $\Omega$ could be the set of all emojis and $\Y$ could be the set of binary vectors of a fixed length. We call $\Omega$ the \emph{universe of (original) objects} and $\Y$ the \emph{space of obfuscated objects}. Intuitively, the algorithm enforces local differential privacy if the probability that an input produces a certain output is roughly equal for all inputs. Formally:
\\[3pt]
\noindent\emph{Let $\A \colon \Omega \to \Y$ be a randomized mechanism. $\A$ satisfies $\epsilon$-local differential privacy if $e^{-\epsilon} \Pr[\A(x') = y] \leq \Pr[\A(x) = y] \leq e^\epsilon \Pr[\A(x') = y]$ for any inputs $x, x'\in \Omega$ and output $y \in \Y$}.
\\[3pt]
We abbreviate the obfuscated object $\A(x)$ with $\x$.

\myparagraph{Count Mean Sketch~\cite{appledifferentialprivacyteamLearningPrivacyScale2017}.}
CMS takes as input objects in the universe $\Omega$ that the user has selected (e.g.\ emojis inserted while typing a message) and returns a binary vector of length $m$ (together with an index), where $m$ is typically much smaller than $|\Omega|$. It uses a family $\H = \{h_1,\ldots,h_{|\H|}\}$ of hash functions that map each object in $\Omega$ to an integer in $\{1,\ldots,m\}$. Given an original object $x \in \Omega$, CMS samples uniformly at random a hash function $h_j \in \H$ and produces the one-hot vector $v_x^{h_j}$ of size $m$ which is 1 at position $h_j(x)$ and 0 in all other entries. The vector $v_x^{h_j}$ can be seen as a compressed version of $x$. Each bit of $v_x^{h_j}$ is then randomly flipped with probability $1 / (1 + e^{\eps/2})$ or left unchanged with the remaining probability $e^{\eps/2} / (1 + e^{\eps/2})$, obtaining the obfuscated vector $\tilde{v}_x^{h_j}$. The output of CMS consists of the obfuscated vector and the index of the hash function used to compute it:
\begin{equation*}
\x = \CMS(x;\ \eps, m, \H) = (\tilde{v}_x^{h_j}, j)
\end{equation*}
CMS satisfies $\eps$-local differential privacy for any $\eps>0$ \cite{appledifferentialprivacyteamLearningPrivacyScale2017}. The parameters $\eps$, $m$ and $\H$ used by CMS on users' devices are typically set by the data curator. In particular, smaller $\eps$ yield lower accuracy, but give better privacy guarantees. Moreover, the hash functions satisfy some technical properties that ensure their behavior is tractable with probabilistic methods --- see Appendix~\ref{sec:CMS} for this and other details on CMS. 

We note that the use of hash functions is not necessary to achieve local differential privacy, but they make CMS more space-efficient and offer additional privacy protection due to hash collisions\footnote{We note that the additional protection coming from collisions is not captured by the privacy loss $\eps$, and hence requires practical attacks like ours to be quantified.}. In fact, even if no bits are flipped, there are often many original objects producing the same one-hot vector, with the exact number depending on $m$ and on the hash function. Collisions make it impossible to infer the original object from the obfuscated object. However, if the user is likely to select most objects from a specific set (pool), after multiple observations this fact can be inferred despite hash collisions. This is the intuition behind our attack.

\section{Pool inference attacks against local differential privacy} \label{sec:theory}

We define a new general attack model against local differential privacy mechanisms, that we call \emph{pool inference attack model}. 
We then propose an attack for this attack model, which we call the Bayesian Pool Inference Attack (BPIA). %

\begin{figure*}[htbp]
\centerline{\includegraphics[width=.95\linewidth]{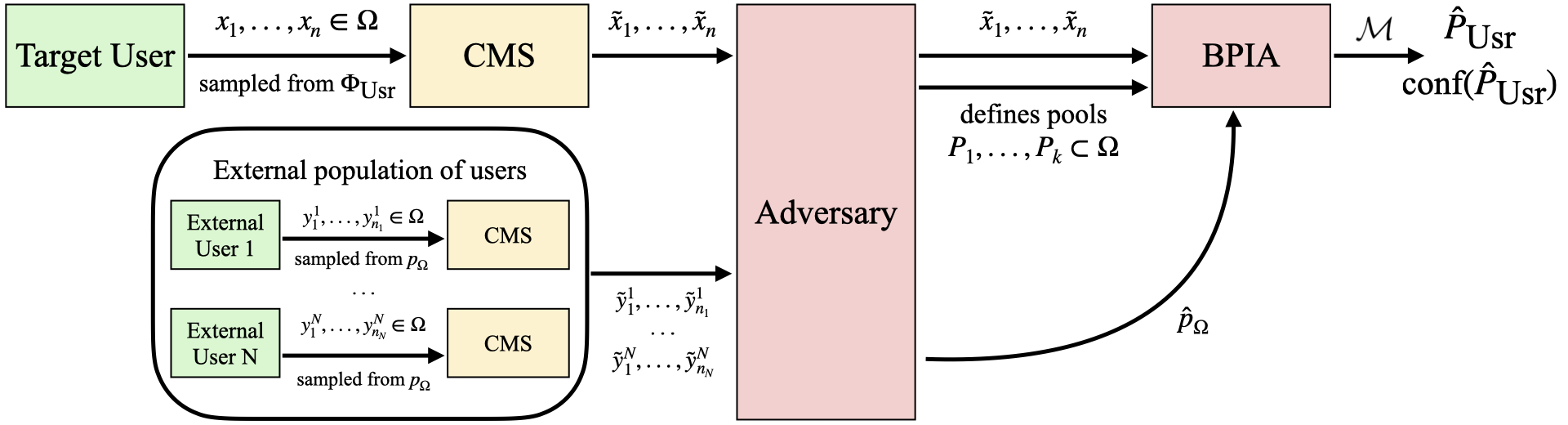}}
\caption{Diagram summarizing the Bayesian Pool Inference Attack (BPIA).}
\label{fig:attack_diagram}
\end{figure*}

\subsection{Formalizing the pool inference attack model}

\begin{table}[t]
\centering
\scriptsize
\begin{tabular}{ r  l l}
\toprule
{\bf Symbol} & {\bf Description} & {\bf Known to $\Adv$} \\ 
\midrule
$\Adv$ & Adversary (runs the attack) & \\
$\U$ & User (target of the attack) & \\
\midrule
$\Omega$ & Universe of (original) objects & \Yes \\
$\Y$ & Space of obfuscated objects & \Yes \\
$\A$ & Mechanism & \Yes \\
$\CMS$ & Count Mean Sketch mechanism & \Yes \\
$\eps$ & Privacy loss parameter & \Yes \\
$\H$ & Family of hash functions & \Yes \\
$m$ & Length of obfuscated vector & \Yes \\
\midrule
$n$ & Number of observations & \Yes \\
$x_1, \ldots, x_n$ & Original objects & \No \\
$\Phi_\U$ & $\U$'s behavior & \No \\
$P_\U$ & $\U$'s preferred pool & \No \\
$\{P_i \colon P_i \neq P_\U\}$ & $\U$'s alternative pools & \No \\
$\gamma_\U$ & $\U$'s relevant interest & \No \\
$\delta_\U$ & $\U$'s polarization & \No \\
$p_\Omega$ & True object popularity & \No \\
\midrule
$\x_1, \ldots, \x_n$ & Obfuscated objects (or observations) & \Yes \\
$P_1, \ldots, P_k$ & $\Adv$'s pools of interest & \Yes \\
$\Omega \sm \cup_{i=1}^k P_i$ & Neutral pool & \Yes \\
$\I$ & $\Adv$'s auxiliary information & \Yes \\
$\score(P_i)$ & $\Adv$'s score for pool $P_i$ & \Yes \\
$\widehat{P}_\U$ & $\Adv$'s estimated preferred pool & \Yes \\
$\conf(\widehat{P}_\U)$ & $\Adv$'s confidence value & \Yes \\
$\ol{\Phi}$ & $\Adv$'s user representation & \Yes \\
$\p_\Omega$ & $\Adv$'s estimated object popularity & \Yes\footnotemark \\

\bottomrule
\end{tabular}
\vspace{-0.1cm}
\caption{Notation and definitions. We indicate which elements are known to the adversary according to the pool inference attack model.}
\label{table:notation}
\vspace{-0.3cm}
\end{table}

\footnotetext{The estimated popularity is always assumed to be known to $\Adv$ but may be uninformative when $\Adv$ uses no auxiliary information, i.e.\ if $\I=\emptyset$.}

We consider an attack where objects are semantically grouped in \textit{pools} (e.g. skin tone of emojis, political orientation of news websites), and the adversary tries to infer which pool a target user samples from most frequently (their \emph{preferred pool}).
Formally, we define the pool inference attack model as a game between an adversary $\Adv$ and a target user $\U$ who obfuscates their data with a mechanism $\A$.
We model the user behavior as a probability distribution $\Phi_\U$ over the universe $\Omega$, reflecting the target user's preferences for the objects in $\Omega$ --- i.e.\ the probability that $\U$ selects a certain object.

\vspace{10pt}
\noindent\fbox{%
\parbox{.97\linewidth}{%
\minipage[t]{\dimexpr\linewidth-2\fboxsep-2\fboxrule\relax}
\myparagraph{Pool Inference Game.}
    \begin{itemize}
\item \emph{Step 1.} $\U$ samples $n$ original objects $x_1,\ldots,x_n$ independently according to $\Phi_\U$. Then, $\U$ runs $\A(x_t)$ independently on each $x_t$, producing the obfuscated objects $\x_1, \ldots, \x_n$.
\item \emph{Step 2.} $\Adv$ selects $k$ \emph{pools of interest} $P_1, \ldots, P_k \sse \Omega$, which are pairwise disjoint subsets of $\Omega$ that can have arbitrary and different sizes.
\item \emph{Step 3.} $\U$ sends $\x_1, \ldots, \x_n$ to $\Adv$.
\item \emph{Step 4.} $\Adv$ runs an attack that returns one pool $\widehat{P}_\U \in \{P_1, \ldots, P_k\}$, which we call $\Adv$\emph{'s estimated preferred pool}.
\end{itemize}
\ \vspace{-15pt}
\endminipage}
}%
\vspace{10pt}
$\Adv$ wins the game if $\widehat{P}_\U = P_{\U}$, where 
\[
P_\U \defeq \argmax\limits_{P_1, \ldots, P_k} \Phi_\U(P_i)
\]
is the user's (true) \emph{preferred pool among} $P_1, \ldots, P_k$.

Without loss of generality, we always assume that the preferred pool $P_\U$ is unique, i.e.\ $\Phi_\U(P_i) < \Phi_\U(P_\U)$ for all $P_i \neq P_\U$. We refer to all the pools in $\{P_i \colon P_i \neq P_\U\}$ as \emph{alternative pools}. We also define the \emph{neutral pool} as the set $\Omega \sm \cup_{i=1}^k P_i$ of all objects not in any pool, and call its elements \emph{neutral objects}. Figure~\ref{fig:skin_tones} provides an illustration of these definitions where the pools are defined in a universe of emojis grouped by skin tone.

We note that, in practice, $\Adv$ could repeat the attack in Step 4 using the same obfuscated objects received in Step 3, but using different pools. For example, the attack could be first run using pools for skin tone, and then again using pools grouped by gender. While this may be a likely case in a real-world setting where the adversary may try to infer as much sensitive information as possible, in this paper we limit our analysis to the case when the attack is run only once for each user (i.e.\ for each instance of the game).

\myparagraph{Adversary's knowledge.}
No information is shared from $\U$ to $\Adv$ or vice versa, except in Step 3, where $\U$ sends the obfuscated objects to $\Adv$. The pools defined by the adversary do not depend on the objects sampled by the user (and vice versa). The only information that $\Adv$ knows about $\U$ are the obfuscated objects $\x_1, \ldots, \x_n$, which we call \emph{observations}. We also admit the possibility that $\Adv$ has access to some auxiliary information $\I$, which represents general knowledge about the population (not about $\U$ specifically) that can be used in the attack in Step 4. Finally, we assume that $\Adv$ knows the universe $\Omega$, the privacy loss $\eps$, and any other internal parameter used when applying $\A$ --- a standard assumption for attacks, where the specifications of the system are assumed to be public. Table~\ref{table:notation} summarizes the notation and what is known to the adversary.

\myparagraph{Behavioral parameters.}
$\U$'s behavior determines how vulnerable they are to a pool inference attack: $\U$ might mostly use objects in the neutral pool, or their preference for their preferred pool might not be strong. For example, $\U$ might use skin-toned emojis only very rarely; moreover, regardless of the relevant interest, it might be that $\U$ selects the medium-light skin tone more frequently, but actually selects other skin tones often as well. To capture these properties of $\U$'s behavior, we define two \emph{behavioral parameters}: the \emph{relevant interest} $\gamma_\U$ (how often $\U$ samples from pools of interest) and the \emph{polarization} $\delta_\U$ (among pools of interest, how often $\U$ samples from their preferred pool). Formally:
\begin{equation*}
\gamma_\U \defeq \Phi_\U \left( \mathsmaller{\bigcup\limits_{i = 1}^k} P_i \right)
\quad \text{and} \quad
\delta_\U \defeq \frac{1}{\gamma_\U} \Phi_\U(P_\U)
\end{equation*}
which satisfy $0 < \gamma_\U \leq 1$ and $\frac{1}{k} < \delta_\U \leq 1$ (since $P_\U$ is maximal with respect to $\Phi_\U$).
While these parameters are unknown to $\Adv$, they are useful to describe each game instance and characterize the user's behavior.
In section~\ref{sec:experiments}, we show that these parameters capture how vulnerable the target user is \emph{with respect to the specific set of pools chosen by $\Adv$}.

\subsection{BPIA: A Bayesian pool inference attack}

We propose an attack using Bayesian inference for the pool inference attack model, that we call BPIA (\emph{Bayesian Pool Inference Attack}). We first summarize the intuition behind the attack. Given $\U$'s obfuscated objects $\x_1,\ldots,\x_n$, BPIA uses Bayesian inference to compute, for each pool $P_i$, the a posteriori probability that $P_i$ is $\U$'s preferred pool:
\begin{equation} \label{eq:posterior_prob}
\Pr[P_\U = P_i \mid \x_1, \ldots, \x_n] 
\end{equation}
To compute this probability, BPIA must take into account (1) the uncertainty on $\U$'s preferred pool and behavioral parameters, (2) the randomness of $\U$'s behavior, and (3) the randomness of the mechanism $\A$. To do this, BPIA uses a hierarchical model that combines the three types of uncertainty. In particular, for (2), BPIA would ideally use the user behavior $\Phi_\U$, but this is unknown to $\Adv$. Instead, BPIA uses a function $\ol{\Phi}$ --- that we call \emph{user representation} --- parameterized by three parameters $\gamma, \delta$ and $\iota$, which models a simple user behavior. We now give the details of the hierarchical model, the user representation and BPIA's output.

\myparagraph{Hierarchical model.}
We propose a general hierarchical model $\M = (\A, \ol{\Phi}, \I)$, where $\A$ is the obfuscation mechanism, $\ol{\Phi}$ is a \emph{user representation} of the (unknown) user behavior,  and $\I$ is some additional auxiliary information that contains general facts about the population (see below).

The user representation is a distribution $\ol{\Phi}(x \mid \iota, \gamma, \delta, \I)$, parameterized by $\iota \in \{1,\dots,k\}$ (the preferred pool), $\gamma \in (0,1]$, $\delta \in (1/k,1]$ (behavioral parameters), and the auxiliary information $\I$. The function $\ol{\Phi}(x \mid \iota,\gamma,\delta,\I)$ gives the (assumed) probability of choosing an original object $x$ if the user has $P_\iota$ as their preferred pool, behavioral parameters $\gamma$ and $\delta$, and subject to additional auxiliary information $\I$.

Intuitively, $\M$ models a user who is first assigned the preferred pool $P_\iota$, the relevant interest $\gamma$ and the polarization $\delta$ uniformly at random; then, the user samples $n$ original objects independently according to $\ol{\Phi}(\cdot \mid \iota, \gamma, \delta, \I)$; and finally obfuscates them using $\A$. Formally, $\M$ is given by three hyperparameters $\iota$, $\gamma$, $\delta$, the random variable $(X_1, \ldots, X_n)$ representing the sampling of the original objects, and the random variable $(\X_1, \ldots, \X_n)$ denoting its randomly obfuscated version, with the auxiliary information $\I$ being treated as a fixed parameter:
\begin{align*}
\iota &\sim \Unif(\{1, \ldots, k\}) \\
\gamma &\sim \Unif((0,1]) \\
\delta &\sim \Unif((1/k,1]) \\
X_t \mid \iota, \gamma, \delta &\sim \ol{\Phi}(\cdot \mid \iota, \gamma, \delta, \I) \quad \forall t \in \{1, \ldots, n\} \\
\X_1, \ldots, \X_n \mid X_1, \ldots, X_n &\sim \A(X_1), \ldots, \A(X_n)
\end{align*}
Using this model, the adversary is able to compute the probabilities in eq.~\ref{eq:posterior_prob}: $\Pr_\M[P_\U = P_i \mid \x_1, \ldots, \x_n]$. While in this model the hyperparameters $\iota$, $\gamma$, and $\delta$ are uniformly distributed --- reflecting an adversary who has no informative prior on $P_\U$, $\gamma_\U$, and $\delta_\U$ --- this could likely be improved in practical settings where the adversary may have access to additional sources of information (see Appendix~\ref{sec:other_aux_info}).

\myparagraph{User representation.}
Our user representation $\ol{\Phi}(x \mid \iota, \gamma, \delta, \p_\Omega)$ models the user assuming the following behavior: the user first chooses a pool (the neutral pool with probability $1-\gamma$, their preferred pool $P_\iota$ with probability $\gamma \delta$, or any of the alternative pools with equal probability $\frac{1}{k-1} \gamma (1-\delta)$), then samples an object from the selected pool according to some \textit{estimated object popularity} $\p_\Omega$.
This object popularity is a distribution over $\Omega$ that --- intuitively --- captures the differences in likelihood for objects \emph{within the same pool}. For example, $\p_\Omega$ can capture the fact that, among emojis with the same skin tone, the thumb-up emoji is much more popular across users than most of the others.
We assume that the adversary has access to this estimated object popularity as additional auxiliary information: $\I = \p_\Omega$. In section~\ref{sec:discussion} we discuss how an adversary could acquire the object popularity from external sources or even estimate it from obfuscated objects collected from other users.
Furthermore, when the adversary does not have any auxiliary information, $\Adv$ can use an \emph{uninformative} object popularity $\p_\Omega$, such as the uniform distribution on $\Omega$.

Formally, the representation $\ol{\Phi}$ is defined as follows:
\begin{equation}
\label{eq:phi_representation}
\ol{\Phi}(x \mid \iota, \gamma, \delta, \p_\Omega) =
\begin{cases}
\gamma \delta \frac{\p_\Omega(x)}{\p_\Omega(P_\iota)} & \text{if } x \in P_\iota \\
\frac{1}{k-1} \gamma (1-\delta) \frac{\p_\Omega(x)}{\p_\Omega(P_i)} & \text{if } x \in P_i, i \neq \iota \\
(1-\gamma) \frac{\p_\Omega(x)}{\p_\Omega(\Omega \sm \cup_{i=1}^k P_i)} & \text{if } x \in \Omega \sm \cup_{i=1}^k P_i
\end{cases}
\end{equation}
We note that in the equation, $\p_\Omega(x)$ is always normalized by the total popularity of the pool that $x$ belongs to. In other words, $\p_\Omega(x)$ is used exclusively to differentiate the probability of different objects within the same pool --- it has no effect on the overall probability that $\ol{\Phi}$ assigns to a pool (and hence to the pool's score, see next paragraph).

We emphasize that the user representation $\ol{\Phi}(x \mid \iota, \gamma, \delta, \p_\Omega)$ is a simple \emph{model} for the user's behavior: $\Adv$ does not know whether the representation correctly describes the actual user behavior $\Phi_\U$, and does not know the exact value of $P_\U$, $\gamma_\U$, and $\delta_\U$.
In particular, our user representation does not account for (1) individual preferences within pools differing from $\p_\Omega$, and (2) preferences between non-preferred pools (since our model assumes that the user selects among alternative pools uniformly at random). In Appendix~\ref{sec:robustness} we present some results that quantify how the correctness of the user representation affects the effectiveness of the attack.

\myparagraph{Maximum a posteriori estimate.} 
The attack attempts to find the user's preferred pool from their obfuscated objects by computing the posterior probability of each pool.
For each pool $P_i$, the adversary computes a \emph{score} proportional to the conditional probability that $P_\U = P_i$ under the model $\M$:
\begin{equation} \label{eq:score_def}
\score(P_i) \propto \Pr_{\M}[P_\U = P_i \mid \x_1, \ldots, \x_n] 
\end{equation}
The adversary then selects the \textit{maximum a posteriori estimate} for the user's preferred pool, as the pool with maximal score:
\begin{equation*} \label{eq:general_estimate}
\widehat{P}_\U = \argmax\limits_{P_1, \ldots, P_k} \score(P_i)
\end{equation*}
If several pools have maximal score, the estimate is selected uniformly at random from these. The attack also computes a confidence value $\conf(\widehat{P}_\U)$ quantifying the probability (under the model $\M$) that the estimate is correct:
\begin{equation*} \label{eq:confidence_estimate}
\conf(\widehat{P}_\U) \stackrel{\text{def}}{=}
\Pr_{\M}[\widehat{P}_\U = P_\U \mid \x_1, \ldots, \x_n] = \frac{\score(\widehat{P}_\U)}{\sum_{i=1}^k \score(P_i)}
\end{equation*}
For an arbitrary confidence threshold $\thr$ defined by the adversary, the attack outputs $\widehat{P}_\U$ if $\conf(\widehat{P}_\U) \geq \thr$ and $\noguess$ otherwise. The threshold $\tau$ hence allows the adversary to set the minimum level of confidence that they require to trust the attack's estimate $\widehat{P}_\U$. The attack is successful if the estimate is correct, i.e.\ $\widehat{P}_\U = P_\U$.

\myparagraph{Score computation.}
Under the model $\M$, the scores defined in eq.~\ref{eq:score_def} are computed as the probability that $P_\U = P_i$ after observing $\x_1,\ldots,\x_n$, obtained by integrating the conditional distribution over $\gamma$ and $\delta$ and applying Bayes's law:
\begin{equation} \label{eq:score_estimate}
\score(P_i) \propto
\int_0^1 \int_{\frac{1}{k}}^1 \prod_{t=1}^n \sum_{z \in \Omega} \Pr_{\A}[\x_t \mid z] \ \ol{\Phi}(z \mid i, \gamma, \delta, \p_\Omega) \ d \delta \ d \gamma
\end{equation}
The term $\Pr_{\A}[\x_t \mid z]$ is the probability that the output of $\A(z)$ is the observation $\x_t$. We give a formal proof of correctness in Appendix~\ref{sec:correctness_score}. We next show how to compute this for CMS.

\myparagraph{Attacking CMS.}
To execute BPIA against the mechanism $\A=\CMS$, we need to determine the value of $\Pr_{\CMS}[\x \mid z]$ for any $\x$ and any $z$. First of all, we note that
\begin{equation*}
\Pr_{\CMS}[\x \mid z] = \Pr_{\CMS}[(\tilde{v}_x^{h_j}, j) \mid z] = \Pr[\tilde{v}_x^{h_j} \mid j, z] \Pr[j \mid z]
\end{equation*}
Since $j$ is selected uniformly at random, we have that $\Pr[j \mid z] = \Pr[j]$ is constant for any $z$ and can be moved outside of the integral in eq.~\ref{eq:score_estimate}. Hence, this is a multiplicative value that is constant across pools and can be ignored.\\
$\Pr[\tilde{v}_x^{h_j} \mid j, z]$ is the probability of obtaining the obfuscated vector $\tilde{v}_x^{h_j}$ when the original object is $z$ and the selected hash function is $h_j$. Since $\Adv$ knows all CMS parameters --- including the hash functions in $\H$ --- they can compute the one-hot vector $v_z^{h_j}$. The probability is then derived by observing how many bits need to be flipped in order to obtain $\tilde{v}_x^{h_j}$ from $v_z^{h_j}$, i.e.\ their Hamming distance. Let $\xi = 1 / (1 + e^{\eps/2})$ be the probability of flipping one bit and let $\lVert \cdot \rVert_1$ denote the $L_1$ norm. We obtain:
\begin{equation} \label{eq:inverted_CMS}
\Pr[\tilde{v}_x^{h_j} \mid j, z] =
\xi^{\bigs\parallel v_z^{h_j} - \tilde{v}_x^{h_j} \bigs\parallel_1} (1-\xi)^{m - \bigs\parallel v_z^{h_j} - \tilde{v}_x^{h_j} \bigs\parallel_1}
\end{equation}
We note that eq.~\ref{eq:inverted_CMS}, when used to compute the score in eq.~\ref{eq:score_estimate}, automatically captures the uncertainty coming from the random flipping of bits and from hash collisions as well. For example, if two objects in different pools share the same hash value, this would make it impossible to distinguish which of them (if any) was $\U$'s original object. The attack takes this fact into account when computing the scores for those pools.

\section{Experiments on synthetic users} \label{sec:experiments}

In this section we empirically validate our BPIA attack against CMS for synthetic users. For each user, we define the behavior $\Phi_\U$ and we then use it to sample the original objects. This allows us to evaluate the attack for different user profiles (relevant interest and polarization) and compare the results across different settings.

\subsection*{Experiment design}

We simulate BPIA in various experiment scenarios. Each \emph{experiment scenario} is defined by the following parameters:

\begin{enumerate}[(i)]
\item the universe $\Omega$;
\item the CMS parameters $\eps$, $m$, $\H$ (see section~\ref{sec:background});
\item the pools of interest $P_1, \ldots, P_k \sse \Omega$ picked by $\Adv$ for the attack;
\item the \emph{true object popularity} $p_\Omega$, a distribution on $\Omega$ (not known to $\Adv$);
\item the \emph{estimated object popularity} $\p_\Omega$ (known to $\Adv$);
\item the number of observations $n$ that $\Adv$ has access to.
\end{enumerate}
Using these parameters, we run 150,000 independent instances of the pool inference game, with one (independent) synthetic user per instance. For each user $\U$, we sample the user's relevant interest $\gamma_\U$ and polarization $\delta_\U$ uniformly at random from $(0,1]$ and $(1/k,1]$, respectively. As will become clear from the results, these behavioral parameters strongly impact the success rate of BPIA. Sampling the parameters uniformly allows us to study the effectiveness of the attack on users with different levels of vulnerability.

\myparagraph{User behavior.}
We select $\U$'s preferred pool $P_\U$ uniformly at random from $\{P_1, \ldots, P_k\}$. For each instance of the game, we use the randomly sampled $\gamma_\U$, $\delta_\U$, and $P_\U$ to define $\U$'s behavior $\Phi_\U$, as follows:
\begin{equation}\label{eq:user_behavior}
\Phi_\U(x) \defeq
\begin{cases}
\gamma_\U \delta_\U \frac{p_\Omega(x)}{p_\Omega(P_\U)} \quad & \text{if } x \in P_\U \\
\frac{1}{k-1} \gamma_\U (1-\delta_\U) \frac{p_\Omega(x)}{p_\Omega(P_i)} \quad & \text{if } x \in P_i \neq P_\U \\
(1-\gamma_\U) \frac{p_\Omega(x)}{p_\Omega(\Omega \sm \cup_{i=1}^k P_i)} \quad & \text{if } x \in \Omega \sm \cup_{i=1}^k P_i
\end{cases}
\end{equation}
This means that to sample each original object, the user first selects $P_\U$ with probability $\gamma_\U \delta_\U$, any other pool of interest with probability $\frac{1}{k-1} \gamma_\U (1-\delta_\U)$ and the neutral pool with probability $1-\gamma_\U$. Once one pool has been selected, the original object is sampled within that pool according to the object popularity $p_\Omega$.

For an instance of the game, we sample $n$ objects from $\Phi_\U$ and obfuscate them with $\CMS({\, \cdot \, }; \eps, m, \H)$. Note here that $\Phi_\U$ corresponds to $\Adv$'s user representation $\ol{\Phi}$ in eq.~\ref{eq:phi_representation} but using $p_\Omega$ (as $\Adv$ does not know the true popularity $p_\Omega$). The robustness results we report in Appendix~\ref{sec:robustness} quantify the effectiveness of the attack when $\U$ uses a noisy version of $p_\Omega$ instead of the exact one.

\myparagraph{Non-private scenario.}
To understand the protection provided by CMS against BPIA, we also report results for an idealized scenario where the mechanism $\A$ simply reveals the original object $x$ (i.e.\ $\A$ is the identity function), and hence $\Adv$ has access to the original objects $x_1, \ldots, x_n$. We refer to this as the \emph{non-private} scenario. In the non-private scenario, BPIA works in the same way as for CMS but the score in eq.~\ref{eq:score_def} is computed by setting $\Pr_\A[\x_t \mid z] = 1$ if $x_t = z$ and $\Pr_\A[\x_t \mid z] = 0$ if $x_t \neq z$.

\myparagraph{Baseline.}
For each scenario, we report as baseline the attack that always makes a guess (i.e.\ has fixed confidence score $\conf=1$) and returns one of the pools $P_1, \ldots, P_k$ uniformly at random. Since we select the user's preferred pool uniformly at random in the experiments, the baseline attack is correct with probability $1/k$.

\myparagraph{Types of adversary.} We simulate two types of adversaries: $\Advw$ and $\Advs$. $\Advs$ has access to auxiliary information on objects' popularity $\p_\Omega$ that approximates $p_\Omega$, while $\Advw$ uses a uniform $\p_\Omega$. 

We consider $\Advs$ to represent a realistic scenario for a typical deployment of local differential privacy (see Discussion). Indeed $\p_\Omega$ can be estimated from auxiliary information derived from an external dataset $\widetilde{\D}_\textit{ext}$ of CMS-obfuscated objects collected from other users. We here simulate $\Advs$'s estimation of $\p_\Omega$ by independently sampling $N = 10^6$ original objects from $p_\Omega$ obtaining $\De = \{y_1, \ldots, y_{N} \}$. We then obfuscate them with CMS, obtaining the external dataset $\widetilde{\D}_\text{ext} = \{ \y_1,\ldots,\y_{N} \}$ which would typically be available to $\Advs$. Using Apple's algorithm~\cite{appledifferentialprivacyteamLearningPrivacyScale2017} the adversary approximates the frequencies of objects of the original dataset $\De$, then projects these frequencies to the probability simplex using alternating projection~\cite{bauschkeProjectionAlgorithmsSolving1996} in order to obtain the estimated object popularity $\p_\Omega$ (which approximates $p_\Omega$ well when the number of users $N$ is sufficiently large).

The estimated object popularity $\p_\Omega$ is the only difference between $\Advw$ and $\Advs$. Both adversaries use the same hierarchical model $\M$ with the same hyperparameters (in particular, they both always integrate over uniformly distributed $\gamma$ and $\delta$ when computing the pools' scores).

Importantly, we note that the effectiveness of the attack in the non-private scenario is the same for $\Advs$ and $\Advw$. In the non-private scenario, there is no uncertainty regarding the original input --- as the output and the input objects coincide --- and hence knowing the object popularity does not bring any advantage.\footnote{This fact can be proved formally by noticing that in the non-private scenario the sum in eq.~\ref{eq:score_estimate} reduces to one single term, so that the object popularity for each observation can be moved outside of the integral and is constant across each pool's score.} 

\myparagraph{Metrics.}
For a given threshold $\tau$, we call \emph{null users} all the users for which the attack does not make a guess ($\conf(\widehat{P}_\U) < \thr$). We then use the following three metrics to measure the effectiveness of our attack in a given scenario:
\begin{enumerate}
\item The \emph{null rate} is the fraction of null users (out of all the 150,000 users) for a given value of $\tau$;
\item The \emph{precision} is the success rate of the attack for all non-null users,  i.e.\ the fraction of non-null users such that $\widehat{P}_\U = P_\U$. That is, the fraction of users for which the attack's guess is correct, out of the users for which a guess is made (which depends on the threshold $\thr$);
\item The \emph{area under the precision-null rate curve} ($\AUC$) is the area under the curve obtained by plotting the precision vs the null rate for all possible threshold values between 0 and 1. Since the threshold $\thr$ can be adapted by the adversary to adjust the tradeoff between precision and null rate, the $\AUC$ captures the overall effectiveness of the attack (in the specific scenario).
\end{enumerate}

\myparagraph{Settings.}
In this paper, we focus on two specific use cases of CMS implemented by Apple in iOS and Mac OS~\cite{appledifferentialprivacyteamLearningPrivacyScale2017}:
\begin{itemize}
\item[] \textbf{Setting 1: Emojis.} In this use case, the device keeps track of which emojis --- the original objects --- are selected by the user when typing. These are obfuscated by CMS and submitted to Apple. The universe of objects contains 2600 emojis, i.e.\ $|\Omega|=2600$.
\item[] \textbf{Setting 2: Web domains.} For this setting, the original objects are the web domains that the user visits using the built-in browser (together with preferences regarding videos autoplay). The implementation of CMS keeps track of 250,000 web domains, i.e.\ $|\Omega|=250000$.
\end{itemize}
Apple's implementation sets $m = 1024$ and $|\H| = 65536$, with $\eps=8$ for web domains and $\eps=4$ for emojis.

\subsection*{Setting 1: Emojis}
\begin{figure*}[htbp]
\centerline{
\includegraphics[width=.85\textwidth]{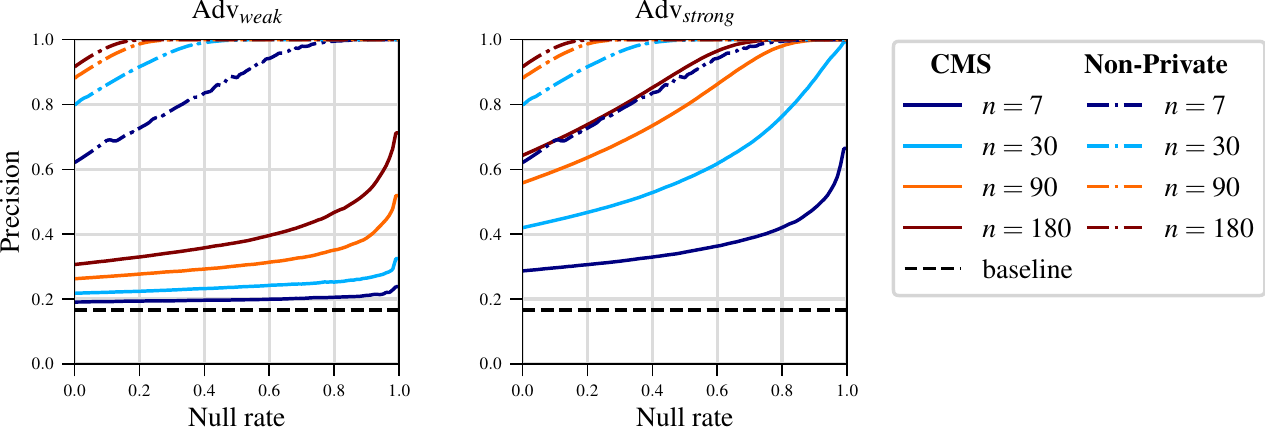}
}
\caption{Precision-null rate curves in the emojis setting for $\Advw$ and $\Advs$. The results for the non-private scenario are the same for $\Advw$ and $\Advs$.}
\label{fig:emojis-prec_vs_null_rate}
\end{figure*}
\begin{figure*}[t]%
\centerline{
\includegraphics[width=.95\textwidth]{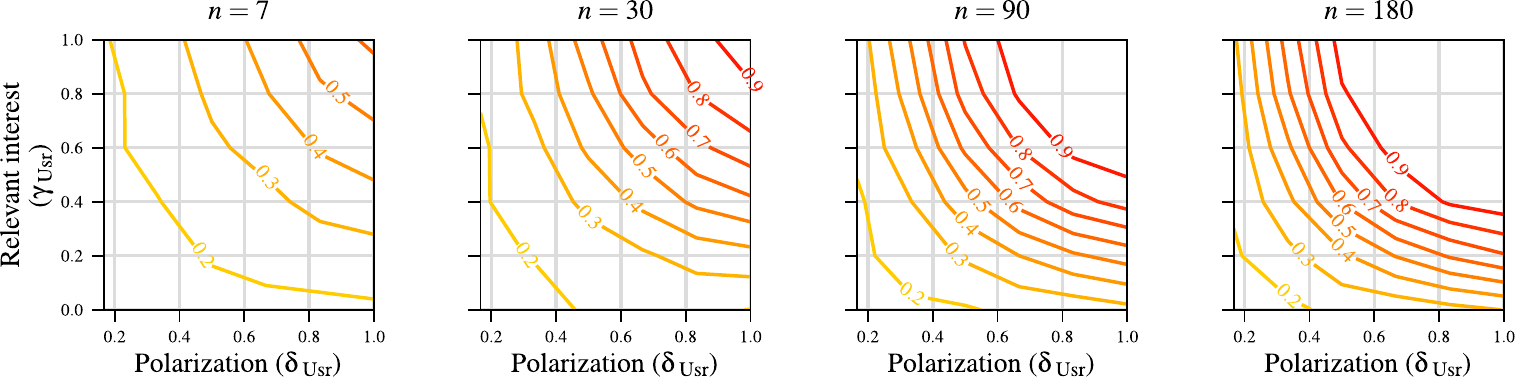}
}
\caption{Precision depending on $\gamma_\U$ and $\delta_\U$ for $\Advs$ in the emojis setting when the attack always makes a guess ($\text{null rate} = 0$). The attack is more efficient when the user's relevant interest and polarization are higher. The figure is generated by computing, for each value of $\gamma_\U$ and $\delta_\U$, the precision of the attack on users with (approximately) those relevant interest and polarization values. We note that $\delta_\U$ is always greater than $1/(6-1) = 0.2$ by definition (see section~\ref{sec:theory}).}
\label{fig:emojis-weak-acc_vs_gamma_and_delta}
\end{figure*}
We consider an adversary $\Adv$ that runs our BPIA attack with the goal of inferring $\U$'s preferred emoji skin tone (see Figure~\ref{fig:skin_tones}). To this end, $\Adv$ defines six pools of size 228, corresponding to the six skin tones supported for 228 emojis in the Unicode Emoji v11.0 standard~\cite{unicodeEmojiListV11}.

We define the true object popularity $p_\Omega$ as a mixture of Zipfian distributions --- reflecting the fact that a few emojis are much more popular than others \cite{erlingssonRAPPORRandomizedAggregatable2014} (we discuss this choice in more detail in Appendix~\ref{sec:zipf_law}). Formally, we consider the partition of $\Omega$ given by the pools $P_1,\ldots,P_k$ and the neutral pool $Q = \Omega \sm \cup_{i=1}^k P_i$. For each $P_i = \{x_1^i,\ldots,x_{|P_i|}^i\}$, we take the Zipfian probability mass function given by:
\begin{equation} \label{eq:zipfs_law}
f_{P_i}(x_j^i)={\frac {1/j^{1.2}}{\sum_{c=1}^{|P_i|} 1/c^{1.2} }}
\end{equation}
and similarly for the neutral pool. Finally we define:
\begin{equation} \label{eq:zipfs_laws_mixture}
p_\Omega(x) \defpropto
\begin{cases}
f_{P_i}(x) \quad & \text{if } x \in P_i, \quad i = 1,\ldots,k \\
f_Q(x)  \quad & \text{if } x \in Q
\end{cases}
\end{equation}

\myparagraph{Results for $\Advw$ and $\Advs$.}
We simulate the attack with $n=7, 30, 90, 180$ observations. Since Apple collects one obfuscated object per day~\cite{appleDifferentialPrivacyOverview}, these correspond to about 1 week, 1 month, 3 months, and 6 months, respectively. While 6 months may seem a long time, most users are likely to keep their iOS and Mac OS devices ---and submit obfuscated objects --- for much longer than that~\cite{statistaSmartphonesReplacementCycle}.

Table~\ref{tab:emojis-AUC-PN} shows that our attack performs well for $\Advs$, already reaching an $\AUC$ of 0.8 after $n=90$ observations. Figure~\ref{fig:emojis-prec_vs_null_rate} shows the attack's full precision-null rate curves. Here again, we see that $\Advs$ performs much better than the baseline, reaching a precision of 0.29 after only 7 observations, and 0.64 after 180 observations when making a guess for all users ($\text{null rate} = 0$). 

\begin{table}[t]%
\begin{center}
\begin{tabular}{ccccc}
\toprule
& $n=7$ & $n=30$ & $n=90$ & $n=180$ \\
\midrule
$\Advw$     &  0.20 &  0.24 &  0.32 &  0.40 \\
$\Advs$     &  0.37 &  0.61 &  0.80 &  0.88 \\
Non-private &  0.86 &  0.96 &  0.99 &  0.99 \\
\bottomrule
\end{tabular}
\end{center}
\caption{$\AUC$ values in the emojis setting.}
\label{tab:emojis-AUC-PN}
\end{table}

Restricting the attack to only users for which the attack is more confident (higher thresholds) allows the adversary to considerably increase the precision while making predictions on a significant number of users. For instance, for $n = 90$, the attack reaches a precision of 1 for a null rate of 0.95. This means that the attack makes no mistake when executed on the top 5\% users (i.e.\ the users whose confidence score is in the top 5\%). Even with a week of observations ($n=7$), the attack reaches 48\% precision (2.9 times better than the baseline) when focusing on the top 10\% of users.

The results for $\Advw$, while significantly better than the baseline, are not as good. When making a guess for all the users, $\Advw$ only reaches a precision of 0.19 for $n=7$ (as opposed to 0.29 for $\Advs$). Even after $n=180$ observations, the precision only increases to 0.31 for $\text{null rate} = 0$, and to 0.53 when focusing on the top 10\% users. These results emphasize the importance of the adversary using auxiliary information during the attack.

The reason $\Advs$ achieves much better results compared to $\Advw$ can be intuitively explained as follows: BPIA uses the object popularity to reduce the indistinguishability of the obfuscated objects. In principle, each obfuscated object may be the output of CMS run on any original object. However, if the attack knows that some of these objects are less likely to be picked (compared to others \emph{in the same pool}), the posterior probability that one of them was the actual input can be reduced accordingly. The score defined in eq.~\ref{eq:score_estimate} captures this fact to compute each pool's posterior probability. We provide additional results on this point in Appendix~\ref{sec:entropy}.

\myparagraph{Results in the non-private scenario.}
In order to contextualize our results, we also measure the accuracy of BPIA in the non-private scenario, when the adversary has access to the user's original objects $x_1, \ldots, x_n$ (i.e. without hashing nor obfuscation).
This gives an upper bound to the attack: even when the adversary observes the original objects, they can still make mistakes when estimating the user's preferred pool.
This is due to the stochastic nature of the user behavior $\Phi_\U$.
For example, a user might use emojis with a certain skin tone most of the times but, for most users, there is a non-zero and possibly significant probability that the user selects emojis with a different skin tone (alternative pool) or even an emoji with no skin tone (neutral pool). Hence, even in the non-private scenario the attack might not be $100\%$ effective.

Figure~\ref{fig:emojis-prec_vs_null_rate} shows the attack to be highly effective in the non-private scenario, although not perfect. 
While the difference in effectiveness between $\Advw$ and non-private remains large for any number of observations, the protection offered by CMS decreases as $n$ increases.

\myparagraph{Impact of the behavioral parameters.}
Figure~\ref{fig:emojis-weak-acc_vs_gamma_and_delta} shows how the precision of the attack increases with both behavioral parameters $\delta_\U$ and $\gamma_\U$ for $\Advs$ when the null rate is 0, i.e. when the attack makes a guess for all users.
Users with larger polarization and relevant interest tend to be, on average, much more vulnerable than other users. For instance for $n=90$ and for $\Advs$, the precision of the attack on a user with $\gamma_\U=0.2$ and $\delta_\U = 0.17$ is lower than 20\%, while it already increases to more than 80\% for a user with $\gamma_\U=0.6$ and $\delta_\U=0.67$. Overall, $\Advs$ performs well over a large range of values of $\gamma_\U$ and $\delta_\U$ for $n \geq 90$.

\subsection*{Setting 2: Web domains}

\begin{figure}[ht]%
\centerline{\includegraphics[width=.8\linewidth]{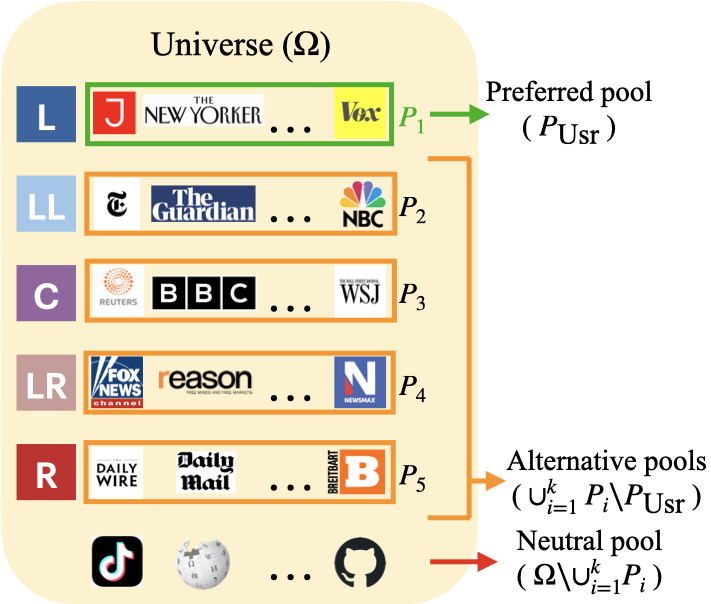}}
\caption{Pools for the web domains setting. Each pool groups together websites for 60 major news outlets according to their political orientation from the 2021 AllSides Media Bias Chart (left, lean left, center, lean right, right). In this case, $\U$ visits most frequently news websites in the left pool.}
\label{fig:allsides_media_bias}
\end{figure}

We here consider the case of an adversary attempting to infer the target user's potential political orientation from news sites that they visit. In this hypothetical setting, the adversary assumes that users are more likely to visit news websites whose political orientation is aligned with their own political views~\cite{iyengarRedMediaBlue2009, stroudNicheNewsPolitics2011}. The adversary hence defines the pools as sets of news websites grouped by political orientation. We here use AllSides's Media Bias rating for 60 major English-language news websites~\cite{allsidesMediaBiasChart2021}. The Chart divides media into five groups: left, lean left, center, lean right, right, which contain respectively 14, 13, 13, 10, and 10 unique news websites\footnote{In a few cases, the chart by AllSides has two entries for the same website --- e.g., for The Wall Street Journal, the \emph{news only} section is rated center and the \emph{opinion} section is rated lean right. As these share the same web domain, for simplicity we include just the \emph{news only} entries in the pools of interest.} (see Figure~\ref{fig:allsides_media_bias}). 
In this experiment we randomly assign a popularity to all websites in the universe. For each object $x \in \Omega$, $p_\Omega(x)$ is sampled uniformly at random from $[0,1]$ (and then rescaled to ensure that $p_\Omega$ has total mass adding up to 1). To reduce the computational time required to run the attack on 150,000 users, we run the experiments with a universe of size $|\Omega| = 2000$ (instead of the original 250{,}000). We show in Appendix~\ref{sec:size_of_universe} that this has no impact on the estimated effectiveness of the attack.

Here again, we simulate two adversaries: $\Advw$ who uses an uninformative (uniform) object popularity $\p_\Omega$, and $\Advs$ who uses $N = 10^6$ obfuscated objects from an external population to derive the estimated popularity $\p_\Omega$.

\myparagraph{Results for $\Advw$ and $\Advs$.}
\begin{figure*}[t]%
\centerline{
\includegraphics[width=.85\textwidth]{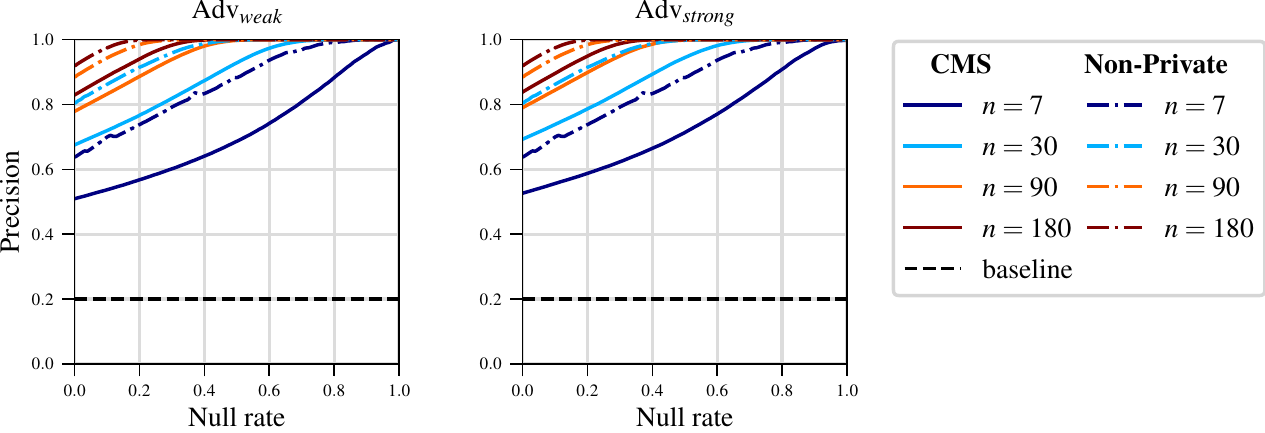}
}
\caption{Precision-null rate curves in the web domains setting for $\Advw$ and $\Advs$. The results for the non-private scenario are the same for $\Advw$ and $\Advs$.}
\label{fig:news-prec_vs_null_rate}
\end{figure*}
\begin{table}[t]%
\begin{center}
\begin{tabular}{ccccc}
\toprule
& $n=7$ & $n=30$ & $n=90$ & $n=180$ \\
\midrule
$\Advw$     &  0.72 &  0.89 &  0.95 &  0.97 \\
$\Advs$     &  0.74 &  0.90 &  0.96 &  0.98 \\
Non-private &  0.87 &  0.96 &  0.99 &  0.99 \\
\bottomrule
\end{tabular}
\end{center}
\caption{$\AUC$ values in the web domains setting.}
\label{tab:news-AUC-PN}
\end{table}
Table~\ref{tab:news-AUC-PN} reports the $\AUC$ of the attack (computed on all users, for any relevant interest and polarization), and shows that both adversaries are very effective. $\Advw$ and $\Advs$ reach high $\AUC$ with few observations. For example, they both obtain $\AUC \geq 0.95$ with $n=90$ observations. 

Interestingly, the effectiveness of $\Advs$ in this scenario is very similar to the one of $\Advw$ --- a stark difference from the emojis setting (Table~\ref{tab:news-AUC-PN} and Figure~\ref{fig:news-prec_vs_null_rate}). This can be explained by the comparatively much smaller pools in this use case compared to the emojis setting (average pool size of 12, compared to 228 in the emoji setting). Indeed as pools get smaller, both the risk of hash collisions between two objects of different pools and the uncertainty introduced by the randomized obfuscation increase (see Appendix~\ref{sec:entropy}).

The small difference in both $\AUC$ and precision, between both adversaries and the non-private scenario further confirms that, when pools are small, CMS provides little additional protection.

\begin{figure*}[htbp]
\centerline{
\includegraphics[width=.95\textwidth]{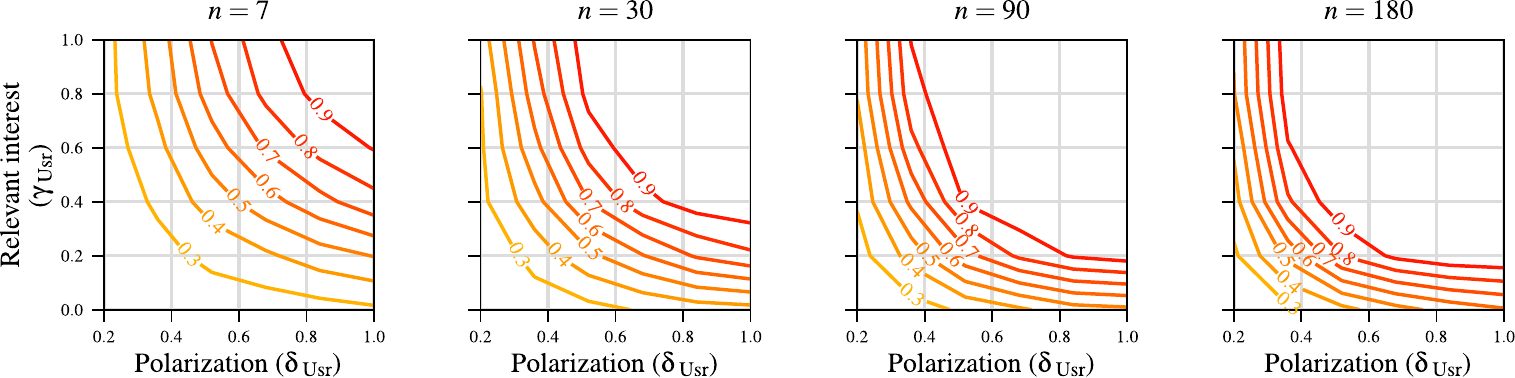}
}
\caption{Precision depending on $\gamma_\U$ and $\delta_\U$ for $\Advs$ in the web domains setting when the attack always makes a guess.}
\label{fig:news-weak-acc_vs_gamma_and_delta}
\end{figure*}

\myparagraph{Impact of the behavioral parameters.}
Figure~\ref{fig:news-weak-acc_vs_gamma_and_delta} shows the precision for $\tau=0$ as a function of the relevant interest $\gamma_\U$ --- the fraction of the time a user visits one of the 60 news websites --- and the polarization $\delta_\U$ for $\Advs$. We omit the results for $\Advw$ as they are almost identical.
Similarly to the emojis setting (Figure~\ref{fig:emojis-weak-acc_vs_gamma_and_delta}), we find that a user's behavioral parameters strongly affect how vulnerable they are.
For instance, for $\Advs$ and $n=90$, the attack will be correct 91\% of the time on a user who visits news websites 20\% of the time ($\gamma_\U=0.2$) and is strongly polarized ($\delta_\U=0.83$) but would only reach 40\% if instead they read diverse sources ($\delta_\U=0.33$) or 25\% if instead they only visit news websites less than 1\% of the time ($\gamma_\U \leq 0.01$).

\myparagraph{Reliability of the confidence score.} 
We have shown that while our attack gives good results overall, it is particularly effective for certain users, in particular users with a high degree of polarization and relevant interest. 

\begin{figure*}[t]%
\centerline{
\includegraphics[width=.95\textwidth]{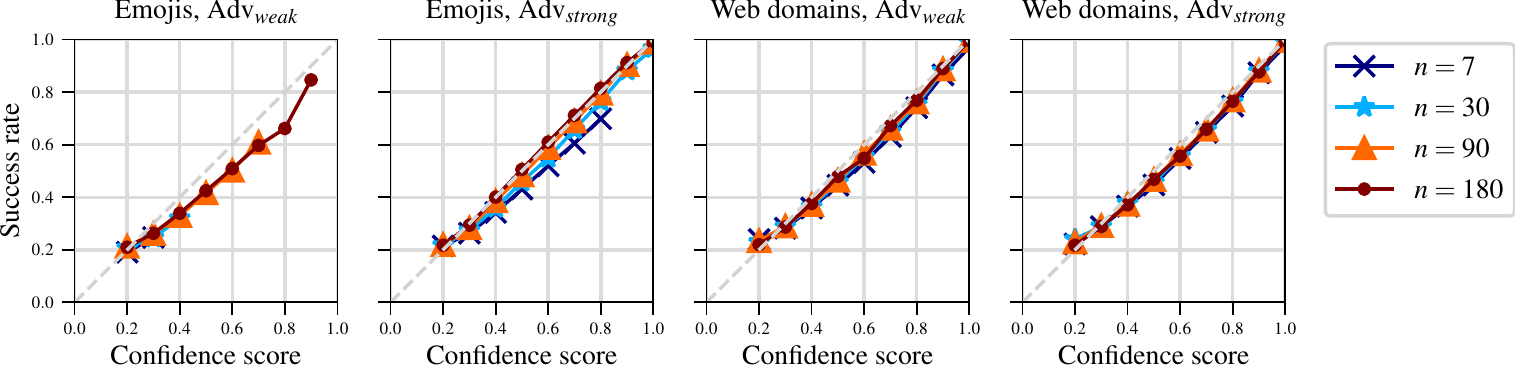}
}
\caption{
Success rate as a function of the confidence score for both $\Advw$ and $\Advs$ and in both the emojis and web domains settings. The confidence score computed by the attack accurately estimates the probability that the attack is correct.
}
\label{fig:all-confidence_score}
\end{figure*}

Figure~\ref{fig:all-confidence_score} shows that our attack's confidence score is well calibrated: for both adversaries, use cases, and number of observations. This makes the attack a concern in practice as it allows an adversary to estimate the probability of the attack to be successful against a specific target user $\U$ by looking only at $\U$'s obfuscated objects.

\section{Experiments on Twitter data} \label{sec:experiments_real_data}

We now simulate the attack in the emojis setting using data collected from Twitter. Our experiments serve two purposes: first, they validate the hierarchical model $\M$ (including the user representation $\ol{\Phi}$) in practice; second, they prove that the attack \emph{can} be very effective on real-world users (see the discussion in section~\ref{sec:discussion}).

\myparagraph{Dataset.}
We use the dataset of tweets collected by Robertson et al.~\cite{robertsonEmojiSkinTone2020}, which contains about 18M tweets from 42K Twitter users, and derive a dataset $\D$ containing only the emojis sent by each users. We then apply a random 80-20 split to $\D$ and obtain: $\Da$, containing the users who used at least one emoji supporting skin tones, on which we simulate the attack; and $\De$, containing the external population that $\Advs$ uses to compute the emojis estimated popularity $\p_\Omega$. We simulate the BPIA attack instantiating the Pool Inference Game on each user in $\Da$, treating each emoji as an original object to which we apply CMS. The full details on how we produce the datasets and run the attack are given in Appendix~\ref{sec:details_twitter_exp}.

\myparagraph{Results for $\Advs$.}
We instantiate the game only for $\Advs$, since our experiments on synthetic data already showed that, in most cases, $\Advw$ is not very effective in the emojis setting.
\begin{figure*}[t]%
\vspace{5pt}
\centerline{
\includegraphics[width=.95\linewidth]{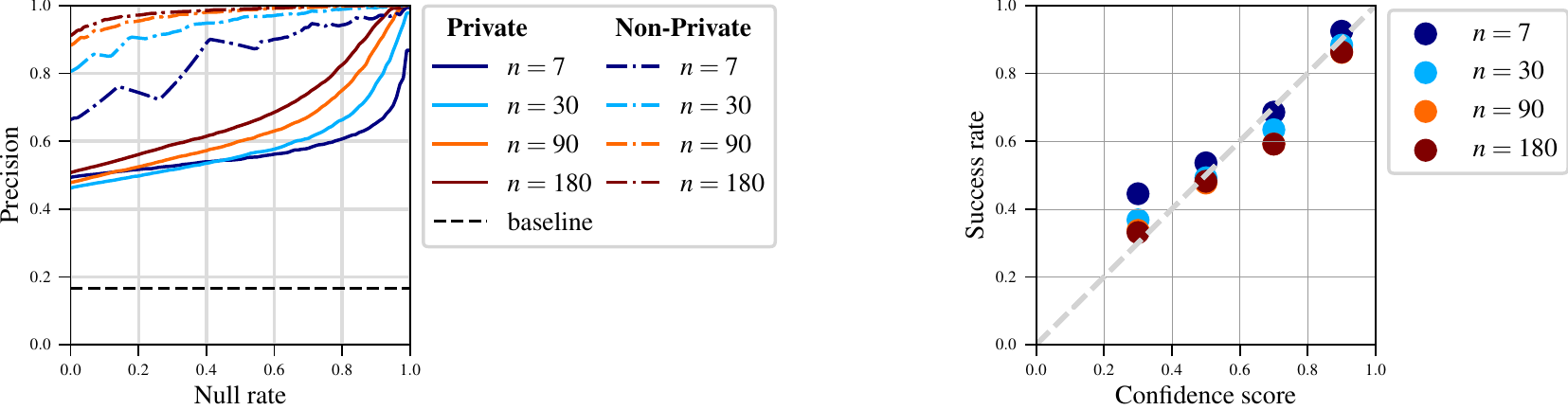}
}
\caption{Precision-null rate curves (\emph{left}) and success rate depending on the confidence score (\emph{right}) for $\Advs$ on Twitter data.}
\label{fig:realdata}
\end{figure*}
Figure~\ref{fig:realdata} (left) shows that the attack is overall very effective on the users in $\Da$. For any number of observations, the precision is $>0.5$ ($2.5$ times better than the baseline) when the attack is run on all the users in $\Da$. After only $n=7$ observations, the precision on the top 20\% of the users (the 20\% of the users with the highest confidence score) is above 0.61, going up to 0.825 after 180 observations. The attack however struggles to reach perfect precision: with $n=180$, to achieve a precision of 0.95 the adversary needs to restrict the attack to the top $10\%$ of the users. This is mostly because, contrary to the synthetic data, $\Da$ contains very few users who have both high $\gamma_\U$ and $\delta_\U$ (see Appendix~\ref{sec:details_twitter_exp}).

Figure~\ref{fig:realdata_cont} shows the precision of BPIA depending on $\gamma_\U$ and $\delta_\U$ when making a guess on every user. These results are mostly consistent with those computed using the synthetic data (Figure~\ref{fig:emojis-weak-acc_vs_gamma_and_delta}). As expected, the attack is not very effective on users with low $\gamma_\U$ and $\delta_\U$, but works remarkably well on high-polarization users who have medium to high relevant interest. For example, after 90 observations, the attack achieves 0.82 to 0.9 precision on the users with polarization over $0.8$ and relevant interest at least $0.4$. Overall, these results validate the applicability of BPIA's model $\M$.
\begin{figure*}[t]%
\centerline{
\includegraphics[width=.95\textwidth]{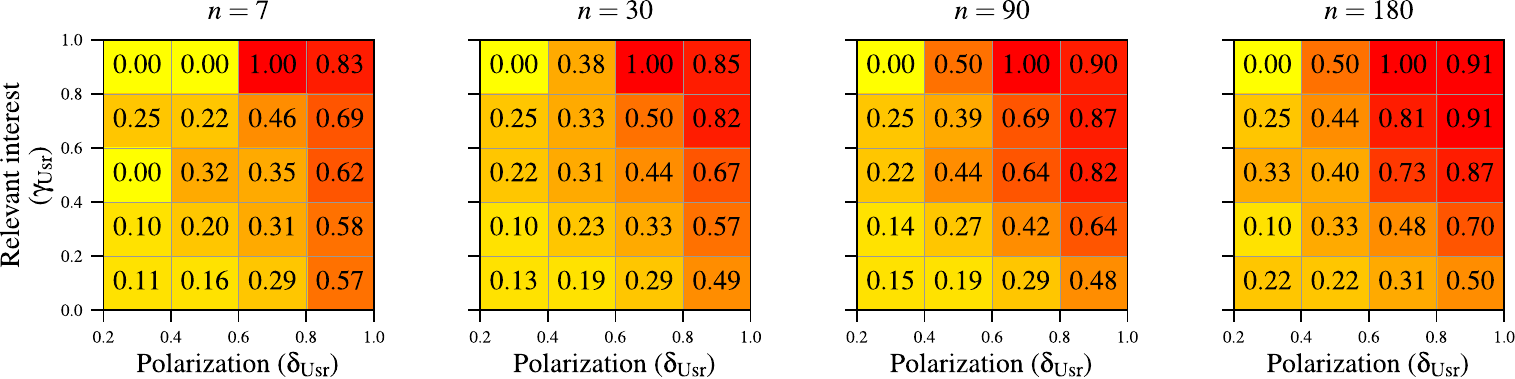}
}
\caption{Precision depending on $\gamma_\U$ and $\delta_\U$ for $\Advs$ on Twitter data when the attack always makes a guess.}
\label{fig:realdata_cont}
\end{figure*}

\myparagraph{Reliability of the confidence score.}
Figure~\ref{fig:realdata} (right) confirms that the confidence score computed by BPIA can be used to accurately estimate the probability that the estimated preferred pool is correct. This validates the fact that BPIA can be used to distinguish and target the most vulnerable users.

\section{Discussion} \label{sec:discussion}

In this paper we propose pool inference, a new attack model that quantifies some practical privacy risks that may affect implementations of local differential privacy mechanisms. We formalize the attack model as a game and propose a Bayesian pool inference attack (BPIA) that applies to any local differential privacy mechanism that processes each object independently. We simulate BPIA against Apple's CMS mechanism for emojis and web domains and study its effectiveness in different scenarios. We show that the attack can successfully allow an adversary to infer sensitive properties of a user's behavior. We further show that BPIA works best on users who are more polarized --- and may hence require the strongest privacy protections. To the best of our knowledge, this is the first attack designed against a real-world implementation of local differential privacy. Taken together, our results show that the BPIA attack is a practical threat for Apple's devices, where CMS is implemented with large $\eps$ parameters and without limiting the cumulative privacy loss after multiple observations.

\myparagraph{Previous criticism to Apple's implementation.}
In September 2017, Tang et al.\ reverse engineered and analyzed Apple's implementation, for which no technical description was yet available. In particular, they found that the choice of the privacy loss $\eps$ was not in line with what is deemed mathematically secure \cite{tangPrivacyLossApple2017}. Tang et al.\ provided a detailed analysis of Apple's system, but did not propose attacks showing how the weakness of the theoretical guarantees could be exploited in practice. Apple disputed the findings by Tang et al., claiming that the system provides far more protection than acknowledged by the researchers~\cite{greenbergHowOneApple}. Our experimental evaluation shows that, with Apple's choice of parameters, our BPIA attack could potentially lead to the disclosure of a user's preference for news websites or emoji skin tone. According to their white paper, Apple discards any user identifier when obfuscated objects are ingested by their servers, making it impossible to later link multiple observations from the same users~\cite{appledifferentialprivacyteamLearningPrivacyScale2017}. While this would limit the attack to a single observation, this is an organizational measure that relies on trust, which is what local differential privacy is designed to avoid~\cite{dworkAlgorithmicFoundationsDifferential2013, josephLocalDifferentialPrivacy2018, xiongComprehensiveSurveyLocal2020} (see also the discussion on mitigation strategies below).

\myparagraph{Representativeness of the experiments.}
In this paper we validate our attack using both synthetically generated data and Twitter data. The goal of our experiments is to study how the attack performs in several scenarios and to validate the user model $\M$ showing that the attack works on a significant number of real-world users. On the other hand, the aim of our experiments is not to measure the fraction of users in the population who are vulnerable. Firstly, while we use Twitter data as a representation of users' usage of emojis, we do not have access to datasets that record such usage across apps or web browsing data. Secondly, our work focuses only on two attack goals (i.e.\ sets of pools): determining the preferred emoji skin tone and the political orientation of the most visited news website. As mentioned in section~\ref{sec:theory}, the adversary can run BPIA as many times as they want \emph{on the same data} using any pools they wish. Users who are not vulnerable to the attack with certain pools might be vulnerable with respect to another set of pools. Moreover, the confidence score can be used to reliably estimate which inferences are likely to be correct. Future work may select other privacy-sensitive pools and use our attack to assess different privacy risks.

\myparagraph{Using auxiliary population-level knowledge to estimate the object popularity.}
Our experiments with $\Advs$ show that an adversary with access to auxiliary information on the overall popularity of objects among the population may be much more effective. The adversary may obtain access to such auxiliary information from a variety of sources, such as social media or studies that report summary statistics on popularity of emojis aggregated over many users. Furthermore, they can be typically estimated by the data curator. In fact, CMS is designed precisely with this scope in mind: estimating the popularity of objects across many users. Hence, if the adversary is the curator themselves, users' privacy is even more at risk.\footnote{We note that assuming that the curator is also the adversary reflects the standard attack model applied to local differential privacy. We believe an external adversary to be less realistic in Apple's case as the obfuscated records are transmitted from the device to Apple through an encrypted connection.} In particular, the method used to estimate the popularity (see section~\ref{sec:theory}) does not require that the adversary knows which objects are collected from which user. $\Advs$ could be a curator who has never acted maliciously before, always discarding the identifiers that would allow to link objects coming from the same user, but who at some point decides to keep together the observations from the same target user.

\myparagraph{Extending the attack to other mechanisms.}
While in this paper we focus on the CMS mechanism, our BPIA attack can be used against any local differential privacy mechanisms where $\Pr_{\A}[\x_t \mid z]$ can be computed analytically or estimated empirically. In Appendix~\ref{sec:HCMS} we show how to adapt the attack to run against HCMS, another mechanism proposed and deployed by Apple to identify websites that cause high usage of hardware resources (CPU and memory)~\cite{appledifferentialprivacyteamLearningPrivacyScale2017}. HCMS is similar to CMS, but uses the Hadamard transform to reduce the size of obfuscated objects to a single bit. Despite this, the way to compute $\Pr_{\HCMS}[\x_t \mid z]$ is similar to the one for CMS.

\myparagraph{Solutions and mitigation strategies.}
There are several possible solutions to protect against BPIA, or at least mitigate it. However, to our knowledge, these all come at a cost in terms of utility, or require significant resources to be deployed.

\noindent \emph{First:} Using a smaller $\eps$ and limiting the total number of observations per user. We show in Appendix~\ref{sec:different_eps} that using a smaller value of $\eps$ reduces the effectiveness of the attack, but it also has a direct impact on utility. Similarly, our results in sections~\ref{sec:experiments} and~\ref{sec:experiments_real_data} show that BPIA is less effective when the number of observed obfuscated objects from the target user is lower, but reducing the total number of observations affects utility as well (see Appendix~\ref{sec:different_eps}). Moreover, limiting the number of observations might make it impossible to learn how users' preferences evolve over time.

\noindent \emph{Second:} Using a local differential privacy mechanism that addresses the privacy loss over multiple observations. These typically use some form of heuristic memoization --- such as Google's RAPPOR~\cite{erlingssonRAPPORRandomizedAggregatable2014} --- or techniques to reduce the number of observations that are collected~\cite{josephLocalDifferentialPrivacy2018}. These may offer a better (theoretical) privacy-utility tradeoff when the population-level distribution that needs to be estimated over time does not change frequently. Extending the pool inference attack model and BPIA to these mechanisms could be used to measure this tradeoff in a practical setting and compare it to the tradeoff provided by CMS.

\noindent \emph{Third:} Adopting a different privacy model. In recent work, researchers have proposed techniques that typically go under the name of \emph{shuffled differential privacy}~\cite{bittauProchloStrongPrivacy2017, cheuDistributedDifferentialPrivacy2019, ballePrivateSummationMultiMessage2020, cheuDifferentialPrivacyShuffle2021, feldmanHidingClonesSimple2021}. This is a hybrid privacy model where the obfuscated objects are routed through an intermediary (the \emph{shuffler}) that in turn sends them to the curator. The role of the intermediary is to shuffle the obfuscated objects to anonymize them and make them unlinkable. Shuffled differential privacy has been deployed by Apple and Google in the context of the Exposure Notification System for COVID-19~\cite{appleExposureNotificationPrivacypreserving2021}. While adopting this model for CMS would protect against BPIA, it effectively moves the requirement of users' trust from the curator to the shuffler: if the two collude, the curator would be able to link the objects again\cite{bittauProchloStrongPrivacy2017, cheuDistributedDifferentialPrivacy2019}. The technical guarantees of the model would be greatly enhanced by using a mix network as the shuffler, but these are extremely hard to deploy in practice~\cite{thedp3tconsortiumDESIREPracticalAssessment2020}. Nevertheless, we believe that the shuffled model is a promising avenue to apply local differential privacy in practice, and we hope this paper will provide evidence of the need for its further development and adoption.

\myparagraph{Source code.}
The code to reproduce the results is available at \url{https://github.com/computationalprivacy/pool-inference}.

\section{Related work}
Our work is part of the line of research studying the guarantees of differential privacy against specific attacks.
Previous research has studied the privacy protections of specific differential privacy mechanisms with respect to attacks that simulate real-world adversaries, but this line of work has so far focused on mechanisms for \emph{central} differential privacy --- the main variant of differential privacy which assumes a trusted curator and one or more untrusted analysts. Examples include attacks against differential privacy mechanisms to release aggregate location time-series~\cite{pyrgelisKnockKnockWho2018, pyrgelisMeasuringMembershipPrivacy2020, pyrgelisWhatDoesCrowd2017}, synthetic data~\cite{stadlerSyntheticDataAnonymisation2021}, and machine learning models~\cite{rahmanMembershipInferenceAttack2018, parkAttackBasedEvaluationMethod2019, jayaramanEvaluatingDifferentiallyPrivate2019}.

To the best of our knowledge, only two papers have empirically investigated the privacy guarantees of a \emph{local} differential privacy mechanism. Pyrgelis et al.~\cite{pyrgelisWhatDoesCrowd2017} propose several attacks on aggregated location data that aim to recover individual users' locations or mobility patterns. They evaluate their attacks against SpotMe~\cite{querciaSpotMEIfYou2011}, a mechanism to obfuscate location data that satisfies local differential privacy~\cite{wasedaAnalyzingRandomizedResponse2016}. Pyrgelis et al.'s work however considers a different adversarial setting than ours: their attacks apply to location time-series obtained by aggregating the obfuscated objects over multiple users, while in our pool inference attack the adversary has access to the individual obfuscated objects. Our attack could be simply adapted to the SpotMe mechanism\footnotemark\ in order to infer the user's preferred pool of locations among some pools of interest --- an interesting application that we leave to future work.
\footnotetext{The SpotMe mechanism is quite similar to CMS, but without hashing. Hence, the probabilities $\Pr_\A[\x \mid z]$ that are used by the attack (eq.~\ref{eq:score_estimate}) can be computed similarly to the ones for CMS.}

Chatzikokolakis et al.~\cite{chatzikokolakisBayesSecurityMeasure2020} propose the Bayes security measure, a general metric that quantifies the expected advantage over random guessing of an adversary that observes the output of a mechanism. They then apply their metric to randomized response~\cite{warnerRandomizedResponseSurvey1965} --- a simple local differential privacy mechanism originally conceived to protect privacy in survey responses. They apply randomized response to the US 1990 Census dataset and find that it gives good protection even for values of $\eps$ as high as 4.8. However, their evaluation focuses on object indistinguishability --- i.e.\ it considers an adversary that collects an obfuscated object and whose goal is to infer the original object. This is a significantly harder objective compared to pool inference and, in fact, CMS's use of hash functions prevents this even for arbitrarily large values of $\eps$. Our work shows that enforcing object indistinguishability is not enough to protect privacy in a practical setting where the adversary has access to multiple obfuscated objects from the same user.

\section{Conclusion}
Apple's implementation of local differential privacy in iOS and Mac OS devices has been presented as a ``technology to help discover the usage patterns of a large number of users without compromising individual privacy''~\cite{greenbergAppleDifferentialPrivacy}. Although researchers have criticized Apple's choice of $\eps$ and unlimited theoretical privacy loss over multiple observations, to our knowledge no practical attacks have been proposed against the mechanisms deployed by Apple. In this paper, we proposed a Bayesian pool inference attack and we empirically evaluated it on Apple's Count Mean Sketch mechanism as configured on Apple's devices. We showed that, especially on the most vulnerable users, the attack could be used to successfully infer (1) the emoji skin tone that the user selects more frequently and (2) the political orientation of the news websites that the user is more likely to visit. Finally, we discussed how the technical privacy guarantees against our attack could be improved, and indicated where further research is necessary to evaluate the privacy/utility tradeoff of these mitigation strategies.

\bibliographystyle{plain}
\bibliography{bibliography}
\appendix

\section{Appendix}

\subsection{Details of CMS} \label{sec:CMS}

The CMS algorithm as proposed by Apple \cite{appledifferentialprivacyteamLearningPrivacyScale2017} is defined by the procedure~\ref{proc:cms}.
\begin{procedure}
    \caption{CMS($x;\ \eps, m, \H$)}\label{proc:cms}
	\KwIn{original object $x$; parameters $\eps$, $m$, $\H$}
	\KwOut{obfuscated object $\x$, index $j$}
	sample $j$ uniformly at random from $\{1,\ldots,|\H|\}$ \;
	$v \gets \{0\}^m$ \;
	$v[h_j(x)] \gets 1$ \;
	sample $b \in \{0,1\}^m$, with $\{b[i]\}_{i=1}^n$ iid and $\Pr[b[i] = 1] = 1 / (1 + e^{\eps/2})$ \;
  \For{$i \gets 0$ \KwTo $m$}{
  \If{$b[i] = 1$}{
    flip $v[i]$
  	  }
	}
	$\vv \gets v$ \;
  \Return $(\vv, j)$
\end{procedure}
The set $\H = \{h_1,\ldots,h_{|\H|}\}$ is a collection of hash functions, where each $h \in \H$ maps every element of $\Omega$ to an integer between 0 and $m-1$. The collection $\H$ is sampled uniformly at random from a family of three-wise independent hash functions~\cite{vadhanPseudorandomness2012}. For any finite sets $A,B$, a family $\mathcal{F}$ of functions $A \to B$ is three-wise independent if, for any mutually distinct $a_1,a_2,a_3 \in A$ and for any $b_1,b_2,b_3 \in B$, we have that $\Pr[f(a_1) = b_1, f(a_2) = b_2, f(a_3) = b_3] = 1/|B|^3$, where the probability is computed over the uniformly random selection of the function $f \in \mathcal{F}$. This property is irrelevant for the differential privacy guarantees, but it contributes to the utility achieved by aggregating CMS objects to estimate frequency histograms~\cite{appledifferentialprivacyteamLearningPrivacyScale2017}. Since Apple does not specify the family used in their implementation of CMS, we generate $|\H|$ fully random hash functions by selecting uniformly at random the value of $h(x)$ for any $h \in \H$ and $x \in \Omega$. This method is not space-efficient, as it requires to store the full description of all functions in $\H$, but it has the advantage of removing any possible source of regularity that might artificially improve the accuracy of our attack. In our implementation, we follow Apple's choice of parameters for the number of hash functions and set $|\H| = 65536$ for all the experiments.

\subsection{Details of the experiments on Twitter data} \label{sec:details_twitter_exp}
\myparagraph{Dataset.}
We use the dataset of tweets collected by Robertson et al.~\cite{robertsonEmojiSkinTone2020}, which consists of about 1.8M tweets collected in 2018 from 42K Twitter users. We consider only users who used emojis at least 10 times across all tweets (approx.\ 26K).

We produce a dataset $\D$ by including, for each user, the first 180 emojis used by the user. This sequence of emojis represents the user's original objects. We then apply a random 80-20 split of $\D$ to obtain $\D_{80}$ and $\D_{20}$. We process $\D_{80}$ to obtain $\Da$ --- containing the users on which we will simulate the attack by $\Advs$ --- and $\D_{20}$ to obtain $\De$, containing the external population that $\Advs$ will use to estimate the emojis popularity $\p_\Omega$:
\\[3pt]
\noindent $\Da$ --- To be able to run the attack with up to $n=180$ observations for each user, if the user has less than 180 objects, we repeat the user's observations in chronological order until we reach 180 objects. For example, if a user originally has only 53 original objects $x_1,\ldots,x_{53}$, we produce 127 new objects such that $x_{54} = x_1, x_{55} = x_2, \ldots, x_{180} = x_{21}$. Finally, we keep only the users who have at least one emoji supporting skin tones, since they are the users on which the attack can be applied ($\gamma_\U > 0$). After this, $\Da$ contains ${\approx} 18$K users with 180 emojis each.
\\[3pt]
\noindent $\De$ --- Similarly, to obtain $\De$ we augment the data in $D_{20}$. This is necessary because $\D_{20}$ contains only 540K original objects in total, while in the experiments with synthetic data the external dataset contains $N = 10^6$ objects (see section~\ref{sec:experiments}). For a fair comparison, we duplicate the original objects in $\D_{20}$ so that the size of $\De$ is $2 \times 540\text{K} \approx 10^6$. We note that this does not affect the validity of our experiments: in a realistic setting, the curator (acting as $\Advs$) would likely have access to millions of obfuscated objects (see section~\ref{sec:discussion}).

\begin{figure}[htbp]
\centerline{\includegraphics[width=.7\linewidth]{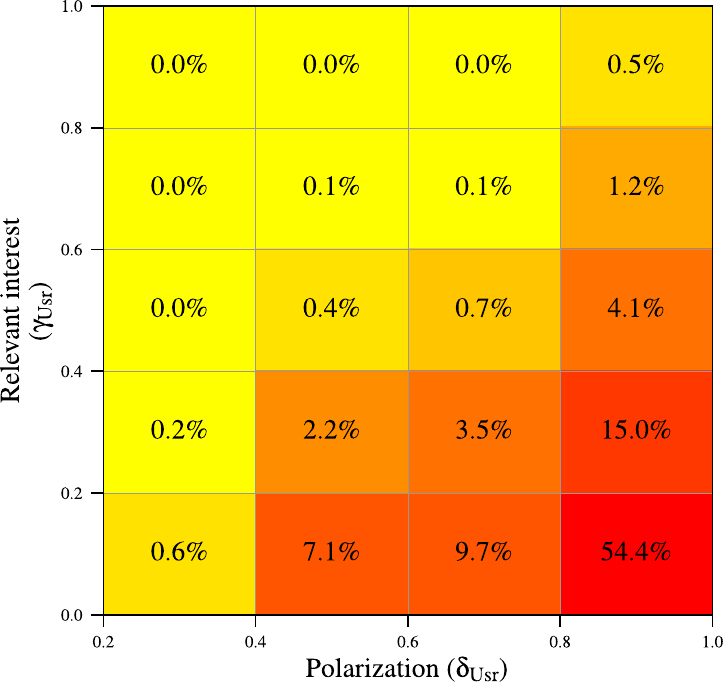}}
\caption{Joint distribution of $(\gamma_\U, \delta_\U)$ in the dataset $\Da$.}
\label{fig:realdata_gammadeltadist}
\end{figure}

\myparagraph{Relevant interest and polarization.}
Using the so obtained dataset $\Da$, we compute each user's preferred pool $P_\U$, relevant interest $\gamma_\U$, and polarization $\delta_\U$ on the full sequence of 180 original objects. We recall that these parameters do not depend on the attack and are not known to $\Advs$, but they are useful to describe how vulnerable $\U$ is. Given $\U$'s sequence of original objects $\underline{x} = x_1, \ldots, x_{180}$ and a subset of the universe $S \sse \Omega$, with an abuse of language we denote $|\underline{x} \cap S| \defeq |\{t \colon x_t \in S\}|$. Then we compute:
\begin{gather*}
P_\U = \argmax_{P_i} |\underline{x} \cap P_i| \\
\gamma_\U = \frac{1}{180} |\underline{x} \cap \mathsmaller{\bigcup\limits_{i = 1}^k} P_i|
\quad \text{and} \quad
\delta_\U = \frac{1}{180\gamma_\U} |\underline{x} \cap P_\U|
\end{gather*}
Figure~\ref{fig:realdata_gammadeltadist} shows how the relevant interest $\gamma_\U$ and the polarization $\delta_\U$ are distributed across users in $\Da$. In particular, the results show that relevant interest is overall not very high, with 71.8\% of the users having $\gamma_\U < 0.2$, meaning that they select emojis supporting skin tone about 20\% of the times or less. On the other hand, most of them have extremely high polarization: $\delta_\U > 0.8$ for 75.2\% of the users.

\myparagraph{Running the attack.}
We instantiate the Pool Inference Game on each of the 18K users in $\Da$, independently. Since we are now using real data, we do not need to define $\U$'s behavior $\Phi_\U$ --- instead, we select the the first $n$ emojis used by $\U$ and run CMS (independently) on them to obtain the $n$ obfuscated records. We note that this choice underestimates the success rate of our attack compared to a random sampling of $n$ objects, since it might be that $\U$'s long-term preferred pool is not the same as in the first $n$ observations on which the attack is run.

\subsection{Effect of entropy} \label{sec:entropy}

For the experiments in section~\ref{sec:experiments}, we have assumed that the object popularity follows a Zipfian distribution (with parameter 1.2) within each pool. While the exact shape of the object popularity is not particularly important, for BPIA to be effective it is important that the pools of interest do not contain a large number of objects with non-negligible popularity. However, this requirement applies only when the pools of interest are large (for example, both $\Advw$ and $\Advs$ achieve good effectiveness in the news case as the pools are small). More precisely, the effectiveness of the attack is lower when the \emph{entropy} of the popularity within pools is higher. Intuitively, this is because large pools ``contain too much noise'', but the noise can be ignored if most of the objects contained in them can be (correctly) ignored by BPIA --- that is, when most of these objects have low \emph{within-pool} probability of being picked by $\U$ (and $\Adv$ knows that).

We now show some results that illustrate this fact more in detail. We consider the emojis setting and we run the attack using six pools of equal size, again assuming that the popularity within each pool (and in the neutral pool) is distributed according to a Zipfian distribution. We vary both the size of the pools ($|P| = 10, 50, 200, 400$) and the Zipfian distribution parameter ($s = 0, 0.5, 1, 2, 4$), for a total of 20 scenarios. Both  these values affect the entropy of the popularity within each pool, as illustrated in Table~\ref{tab:entropy_values}. We note that the parameter $s = 0$ yields a uniform distribution.
\begin{table}[t]
\begin{center}
\begin{tabular}{l|ccccc}
\toprule
$|P|$ & $s = 0$ & $s = 0.5$ & $s = 1$ & $s = 2$ & $s = 4$ \\
\midrule
10  & 3.32 &  3.22 &  2.88 &  1.78 &  0.48 \\
50  & 5.64 &  5.44 &  4.61 &  2.19 &  0.48 \\
200 & 7.64 &  7.36 &  5.99 &  2.31 &  0.48 \\
400 & 8.64 &  8.33 &  6.64 &  2.33 &  0.48 \\
\bottomrule
\end{tabular}
\end{center}
\caption{Entropy of the popularity within each pool of interest when the pool has size $|P|$ and the distribution within the pool follows Zipfian distribution with parameter $s$.}
\label{tab:entropy_values}
\end{table}
For simplicity, we simulate an adversary that knows the true popularity $p_\Omega$ and uses it in BPIA, i.e.\ $\Adv$ sets $\p_\Omega = p_\Omega$. We run BPIA against 1000 users in each scenario, again using eq.~\ref{eq:user_behavior} to define the user behavior.

Table~\ref{tab:entropy_exp_AUC-PNs} shows that the AUC-PN of the attack is negatively affected by the entropy of the popularity within pools. For any size of the pools, larger values of $s$ lead to lower entropy (Table~\ref{tab:entropy_values}), which results in better effectiveness of the attack. When the entropy is very low (e.g.\ for $s=4$), the attack is very effective even when the pools contain 400 objects ($\AUC=0.98$). On the other hand, when $s=0$ --- so that the popularity is the uniform distribution ---, the entropy is maximal, but the AUC-PN is significantly affected only for larger pools.

\begin{table}[ht]
\begin{center}
\begin{tabular}{l|ccccc}
\toprule
$|P|$ & $s = 0$ & $s = 0.5$ & $s = 1$ & $s = 2$ & $s = 4$ \\
\midrule
10  &  0.89 &  0.90 &  0.92 &  0.96 &  0.98 \\
50  &  0.69 &  0.73 &  0.85 &  0.97 &  0.98 \\
200 &  0.43 &  0.52 &  0.80 &  0.96 &  0.98 \\
400 &  0.35 &  0.42 &  0.78 &  0.96 &  0.98 \\
\bottomrule
\end{tabular}
\end{center}
\caption{AUC-PN of the attack for $n=180$ observations depending on the size and distribution within the pools.}
\label{tab:entropy_exp_AUC-PNs}
\end{table}

\subsection{Robustness of the attack} \label{sec:robustness}

In our experiment design, the user's behavior $\Phi_\U$ is defined in eq.~\ref{eq:user_behavior} using the exact object popularity $p_\Omega$ for all users. This is equivalent to assuming that while users may have different preferences regarding the pools (as determined by their $\gamma_\U$ and $\delta_\U$), their preferences \emph{within} the pools are always distributed according to $p_\Omega$. In particular, this provides an advantage for $\Advs$, who has access to the estimated object popularity $\p_\Omega$ which approximates $p_\Omega$. However, in practice it is unlikely that $\U$'s preferences within pools follow \emph{exactly} the population-level object popularity.

We now show how the success rate of the attack for $\Advs$ degrades as $\U$'s preferences within the pools become more and more different from the object popularity $p_\Omega$. We run again the experiments for the emojis setting, but we now define $\U$'s behavior using a randomly perturbed object popularity. Formally, we add Gaussian noise to each object's popularity, independently:
\begin{equation*}
\tilde{p}_\U(x) \sim p_\Omega(x) + \mathcal{N}(0,\sigma^2) \quad \forall x \in \Omega
\end{equation*}
Since $\tilde{p}_\Omega$ might not be a probability distribution, we turn it into one by ensuring that it takes only positive values and the total mass is 1:
\begin{equation*}
p_\U(x) \defeq 
\frac{\tilde{p}_\Omega(x) + |\min_{y \in \Omega} p(y)|}
{\sum_{z \in \Omega} (\tilde{p}_\Omega(z) + |\min_{y \in \Omega} p(y)|)}
\end{equation*}

We then define the user behavior as we did before, but using the user-specific perturbed popularity $p_\U$:
\begin{equation*}\label{eq:user_behavior_noisy}
\Phi_\U(x) =
\begin{cases}
\gamma_\U \delta_\U \frac{p_\U(x)}{p_\U(P_\U)} \quad & \text{if } x \in P_\U \\
\gamma_\U (1-\delta_\U) \frac{p_\U(x)}{p_\U(\cup_{i=1}^k P_i \sm P_\U)} \quad & \text{if } x \in P_i \neq P_\U \\
(1-\gamma_\U) \frac{p_\U(x)}{p_\U(U \sm \cup_{i=1}^k P_i)} \quad & \text{if } x \in U \sm \cup_{i=1}^k P_i
\end{cases}
\end{equation*}
We note that the larger the standard deviation of the noise $\sigma$, the larger the difference between $p_\U(x)$ and the object popularity $p_\Omega(x)$, and hence (most likely) between $p_\U(x)$ and $\Adv$'s estimated object popularity $\p_\Omega(x)$ as well.

We simulate the attack in the emojis setting for $\sigma = 10^{-5}, 10^{-4}, 10^{-3}, 10^{-2}$. The case $\sigma = 0$ corresponds to the original scenario for $\Advs$ in section~\ref{sec:experiments}.

To quantify the distance between the estimated popularity $\p_\Omega$ and the noisy version $p_\U$, we compute the Jensen-Shannon divergence between the two distributions, defined as follows:
\begin{equation*}
\operatorname{JSD}(\p_\Omega, p_\U) = \frac{1}{2} D_{\rm KL}(\p_\Omega \parallel M) + \frac{1}{2} D_{\rm KL}(p_\U \parallel M)
\end{equation*}
where $M = \frac{1}{2}(\p_\Omega + p_\U)$ and $D_{\rm KL}(\cdot \parallel \cdot)$ is the Kullback–Leibler divergence. The Jensen-Shannon divergence
is bounded between 0 and 1.

We then compute the average of these distances across users:
\begin{equation*}
\avgd \defeq \frac{1}{n_\text{users}} \sum_{\U} \operatorname{JSD}(\p_\Omega, p_\U)
\end{equation*}
where $n_\text{users} = 1000$, as these are enough to accurately estimate the attack's AUC-PN.

Table~\ref{table:robustness_AUC-PN_advs} shows that $\Advs$'s effectiveness gets worse as the magnitude of the noise gets larger. However, the decrease in effectiveness is rather limited when $\avgd$ is not too large. For example, for $\sigma = 10^{-4}$, the AUC-PN after 180 observations is 0.79, just 0.09 points less than for $\sigma=0$. Hence, in the emojis setting BPIA preserves its effectivess when $\U$'s preferences within pools do not deviate too much from the estimated popularity $\p_\Omega$.

\begin{table}
\centering
\setlength\tabcolsep{3pt} %
\begin{tabular}{c|c|cccc}
\toprule
$\sigma$ & $\avgd$ & $n=7$ & $n=30$ & $n=90$ & $n=180$ \\
\midrule
0          &  0.05 &       0.36 &        0.60 &        0.79 &         0.88 \\
$10^{-5}$ &  0.07 &       0.36 &        0.60 &        0.78 &         0.87 \\
$10^{-4}$ &  0.24 &       0.30 &        0.47 &        0.68 &         0.79 \\
$10^{-3}$ &  0.60 &       0.20 &        0.24 &        0.30 &         0.37 \\
$10^{-2}$ &  0.75 &       0.18 &        0.19 &        0.21 &         0.22 \\
\bottomrule
\end{tabular}
\caption{Effectiveness of the attack for $\Advs$ in the emojis setting depending on the level $\sigma$ of the noise applied to obtain the user's distribution $p_\U$. The four columns on the right report the AUC-PN.}
\label{table:robustness_AUC-PN_advs}
\end{table}

We run the same experiments for the web domains setting, but we omit the results as the the effectiveness of the attack is almost identical independently of the noise. This is to be expected, as in the web domains setting the effectiveness for $\Advs$ is almost identical to the one for $\Advw$, who has no auxiliary information and uses a uniform estimated popularity $\p_\Omega$.

\subsection{Size of the universe} \label{sec:size_of_universe}

Apple's implementation of CMS for the web domains setting uses a universe $\Omega$ containing 250,000 objects. In order to reduce the computational time required to run BPIA on 150,000 users, for the experiments in section~\ref{sec:experiments} we use a smaller universe containing 2,000 objects. We now show that the size of the universe (and, in particular, of the neutral pool) has close to no impact on the effectiveness of the attack. 

We run BPIA in the exact same scenario, changing only the value of $|\Omega|$ --- in particular, we keep the same pools of size 14, 13, 13, 10, and 10. We use universe sizes $|\Omega| = 1000, 10000, 100000, 250000$. These values are used both in the simulation of the users and in the simulation of the adversary. For each size, we run the simulation on 5,000 users --- which leads to a sufficiently accurate estimate of the AUC-PN.

Table~\ref{table:size_of_universe} reports the AUC-PN values for $\Advw$, which are almost identical for all universe sizes and across all number of observations. We omit the results for $\Advs$ as they are very similar. Intuitively, the size of the universe is mostly irrelevant to BPIA's effectiveness because the only relevant bits in the obfuscated objects are the ones associated (by the randomly selected hash function) with an original object that belongs to a pool of interest. Hence, objects from the neutral pool are relevant only if they yield a collision with any of these. Since the hash function is randomly selected, the collisions tend to distribute evenly inside the pools across multiple observations.

\begin{table}
\centering
\begin{tabular}{l|cccc}
\toprule
$|\Omega|$ & $n=7$ & $n=30$ & $n=90$ & $n=180$ \\
\midrule
   1000 &       0.72 &        0.90 &        0.96 &         0.97 \\
  10000 &       0.71 &        0.88 &        0.95 &         0.97 \\
 100000 &       0.72 &        0.89 &        0.95 &         0.98 \\
 250000 &       0.72 &        0.89 &        0.95 &         0.97 \\
\bottomrule
\end{tabular}
\caption{Effectiveness of the attack for $\Advw$ in the web domains setting depending on the size of the universe. The four columns on the right report the AUC-PN.}
\label{table:size_of_universe}
\end{table}

\subsection{Effect of $\eps$} \label{sec:different_eps}

\begin{table*}
\centering
\begin{tabular}{l|rrrr|rrrr}
\multicolumn{2}{c}{} & \multicolumn{2}{c}{$\MAE(\dot{p}_\Omega, p_\Omega)$} & \multicolumn{2}{c}{} & \multicolumn{2}{c}{$\MAPE(\dot{p}_\Omega, p_\Omega)$} \\
\midrule
$\eps$ &  $Z=10^6$ & $Z=10^7$ & $Z=10^8$ & $Z=10^9$ &  $Z=10^6$ & $Z=10^7$ & $Z=10^8$ & $Z=10^9$ \\
\midrule
0.01 & 0.163794 & 0.050822 &  0.015939 &   0.005230 & 184449.89\% & 55912.08\% & 17980.67\% &   5835.21\% \\
0.10 & 0.016129 & 0.005109 &  0.001553 &   0.000493 &  18304.38\% &  5665.86\% &  1849.65\% &    561.66\% \\
1 & 0.001572 & 0.000506 &  0.000156 &   0.000052 &   1659.39\% &   563.36\% &   174.64\% &     57.87\% \\
4 & 0.000337 & 0.000111 &  0.000036 &   0.000016 &    376.18\% &   119.29\% &    40.34\% &      18.50\% \\
8 & 0.000115 & 0.000038 &  0.000016 &   0.000013 &    126.35\% &    40.42\% &    18.55\% &     14.41\% \\
\bottomrule
\end{tabular}
\caption{Utility achieved by the curator to estimate the popularity distribution $p_\Omega$ when CMS is run using privacy loss $\eps$ and $Z$ CMS-obfuscated objects are collected.}
\label{table:news_utility}
\end{table*}

The parameter $\eps$ controls the level of the noise in CMS, i.e.\ the probability that each bit is flipped. The value of $\eps$ hence affects the privacy guarantees and, in turn, the effectiveness of BPIA. To quantify this impact, we simulate the attack in the web domains setting, changing only the value of $\eps$. We then show how using smaller $\eps$ affects utility.

\myparagraph{Effect on the attack.}
Table~\ref{table:news_eps} reports the results for $\Advw$. As expected, the value of $\eps$ heavily impacts the AUC-PN. Interestingly, for $\eps=0.1$ the AUC-PN does not improve over the baseline of $0.2$ even after $n=180$ observations --- when the total (theoretical) privacy loss is $\eps_\text{tot} = 180 \times 0.1 = 18$. This is remarkable because when $\eps_\text{tot} = 18$, there are virtually no theoretical privacy guarantees, and yet such $\eps$ is sufficient to fully protect against BPIA in the web domains setting. This highlights the importance of quantifying the privacy guarantees of differential privacy mechanisms against realistic attack models and scenarios when the theoretical privacy loss is large~\cite{dworkDifferentialPrivacyPractice2019}.

\begin{table}
\centering
\begin{tabular}{l|rrrr}
\toprule
  $\eps$ &  $n=7$ &  $n=30$ &  $n=90$ &  $n=180$ \\
\midrule
 0.01 &       0.20 &        0.20 &        0.20 &         0.20 \\
  0.1 &       0.20 &        0.20 &        0.20 &         0.20 \\
    1 &       0.23 &        0.29 &        0.36 &         0.40 \\
    4 &       0.40 &        0.63 &        0.81 &         0.88 \\
    8 &       0.72 &        0.90 &        0.96 &         0.97 \\
\bottomrule
\end{tabular}
\caption{Effectiveness of the attack for $\Advw$ in the web domains setting depending on the value of $\eps$. The four columns on the right report the AUC-PN.}
\label{table:news_eps}
\end{table}

\myparagraph{Effect on utility.} 
The results in Table~\ref{table:news_eps} show that BPIA could be made significantly less effective --- at least in our setting --- by using $\eps \leq 1$ in CMS. We now show that this would however seriously impact the utility of the CMS-obfuscated objects that are collected and aggregated by the curator. To show this, we measure the accuracy of the object popularity that would be estimated by the data curator, under the different $\eps$ values. Since this accuracy also heavily depends on the number of CMS-obfuscated objects that are collected, we show the results for different numbers of CMS-obfuscated objects $Z = 10^6, 10^7, 10^8, 10^9$. This can be interpreted as the total number of objects by all users. For example, if users send 100 objects each on average, then $10^7$ users are necessary to collect $10^9$ objects.

For a given value of $\eps$ and $Z$, we draw $Z$ original objects from the universe according to the true popularity $p_\Omega$. We then apply CMS with the given $\eps$ to obtain $Z$ obfuscated objects, and use Apple's algorithm to derive an estimation of the popularity $\dot{p}_\Omega$. Finally, we measure the accuracy of $\dot{p}_\Omega$ by using two metrics: the mean absolute error $\MAE(\dot{p}_\Omega, p_\Omega)$ and the mean absolute percentage error computed on the top 80\% objects\footnotemark\ $\MAPE(\dot{p}_\Omega, p_\Omega)$.
\footnotetext{Since the MAPE is very sensitive to error on objects with very small probabilities, ignoring the 20\% of objects with the lowest probability gives a clearer measure of utility. In practical deployments of local differential privacy, determining the exact popularity of unpopular objects is likely not necessary.}

The results in Table~\ref{table:news_utility} show that the accuracy of $\dot{p}_\Omega$ are greatly affected by $\eps$, with both the MAE and the MAPE increasing about linearly with $\eps$ for any value of $Z$. In particular, the results show that $\dot{p}_\Omega$ starts reaching an acceptable accuracy ($\MAPE = 18.55\%$) only when $\eps=8$ and $10^8$ CMS objects are collected. Using $\eps=1$ for the same number of objects would result in a much larger error ($\MAPE = 174.64\%$). Even when the curator collects $10^9$ CMS-obfuscated objects using $\eps=1$ still results in much lower utility: the MAPE is $57.87\%$, about three times as much as for $\eps=8$ and $Z=10^8$. While an extensive analysis of the utility of CMS for different use cases is beyond the scope of this paper, these results suggest that mitigating our attack by using a smaller $\eps$ parameter in CMS would likely affect utility to an unacceptable level.

\subsection{Choice of the object popularity} \label{sec:zipf_law}

\begin{figure*}[t]
\centerline{
\includegraphics[width=.8\textwidth]{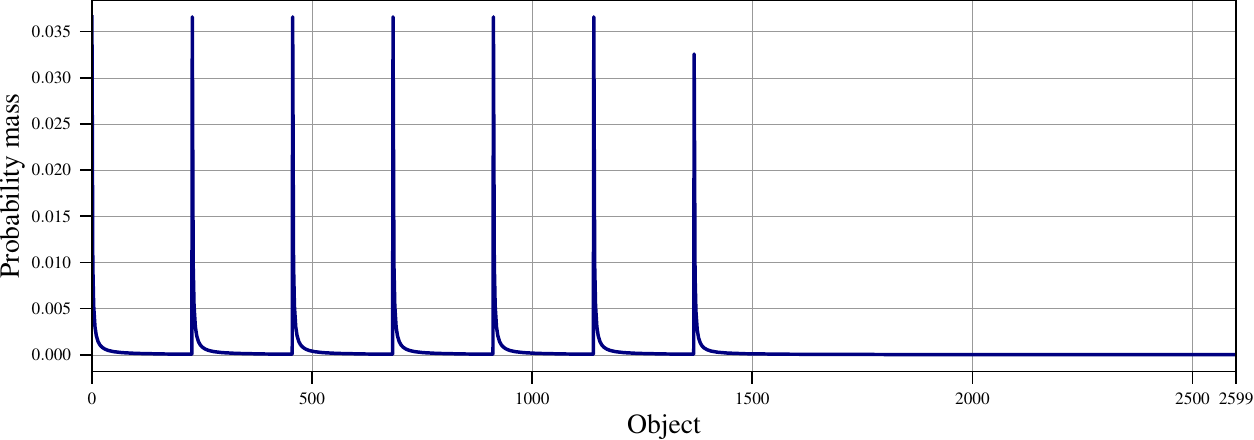}
}
\caption{
Probability density function corresponding to the true object popularity $p_\Omega$ used in the experiments for the emojis setting. The function is obtain as a mixture of seven Zipfian distributions: six for the pools of interest and one for the neutral pool.
}
\label{fig:mixture_zipf_law}
\end{figure*}

In the experiments for the emojis setting, we define the true object popularity as a mixture of Zipfian distributions --- see eq.~\ref{eq:zipfs_law} and \ref{eq:zipfs_laws_mixture}. Figure~\ref{fig:mixture_zipf_law} shows the resulting distribution.

Zipfian distributions are often used as the underlying popularity distribution in the context of frequency estimation with local differential privacy \cite{erlingssonRAPPORRandomizedAggregatable2014, appledifferentialprivacyteamLearningPrivacyScale2017, wangLocallyDifferentiallyPrivate2017}. Moreover, empirical work on Twitter data has found that the popularity of emojis is approximated by a power law with an exponential cutoff~\cite{novakSentimentEmojis2015}, which is similar to a Zipfian distribution in that most of the probability mass is concentrated on few objects.

In choosing to use a mixture of Zipfian distributions --- rather than just one Zipfian distribution --- we are implicitly assuming that the distribution is the same across skin tones. While this is a working assumption that might not accurately reflect the real-world usage of emojis, it would likely not affect the effectiveness of BPIA in a real-world setting. As we showed in Appendix~\ref{sec:entropy}, the effectiveness of BPIA in the emojis setting for $\Advs$ depends on the fact that --- for each skin tone --- a few emojis account for most of the probability mass. Since this fact holds true for emojis overall, we believe that it holds for each skin tone as well.

\subsection{HCMS} \label{sec:HCMS}

Hadamard Count Mean Sketch (HCMS) is another local differential privacy mechanism proposed by Apple~\cite{appledifferentialprivacyteamLearningPrivacyScale2017}. It is a variant of CMS where every obfuscated object consists of only one bit. This is achieved by applying the Hadamard basis transform to the one-hot vector $v_x^{h_j}$ and sampling only one bit from the transformed vector that is then obfuscated and submitted. Given $s \in \N$, the Hadamard basis $H_s$ is the $2^s \times 2^s$ matrix such that $H_0 = 1$ and:
\begin{equation*}
H_s = 
\begin{pmatrix}
H_{s-1} & H_{s-1} \\ H_{s-1} & -H_{s-1}
\end{pmatrix}
\end{equation*}
The HCMS mechanism is described in procedure~\ref{proc:hcms}.
\begin{procedure}
    \caption{HCMS($x;\ \eps, m, \H$)}\label{proc:hcms}
	\KwIn{original object $x$; parameters $\eps$, $m$, $\H$}
	\KwOut{obfuscated object $\x$, index $j$}
	sample $j$ uniformly at random from $\{1,\ldots,|\H|\}$ \;
	$v \gets \{0\}^m$ \;
	$v[h_j(x)] \gets 1$ \;
	$w \gets H_s v$, where $s \gets \log_2 m$ \;
	sample $l$ uniformly at random from $\{0,\ldots,m-1\}$ \;
	sample $b \in \{0,1\}$, with and $\Pr[b = 1] = 1 / (1 + e^{\eps})$ \;
  \If{$b = 1$}{
    flip $v[l]$
  	  }
	$\vv \gets v[l]$ \;
  \Return $(\vv, j, l)$
\end{procedure}
Apple uses HCMS on a universe $\Omega$ of 250,000 web domains, with parameters $m = 32768$, $|\H| = 1024$, and $\eps = 4$.

Our BPIA attack can be extended to HCMS by adapting the term $\Pr_{\A}[\x_t \mid z]$ in eq.~\ref{eq:score_estimate}, i.e.\ by computing $\Pr_{\HCMS}[(\vv_t, j_t, l_t) \mid z]$. This is done by applying the Hadamard transform to $v_z^{j_t}$ and observing if the $l_t$th bit has been flipped. Let $w_z^{j_t} = H_s v_z^{j_t}$ with $s = \log_2 m$ be the transformed vector and let $\xi' = 1 / (1 + e^{\eps})$ be the probability of flipping the selected bit. Then:
\[
\Pr_{\HCMS}[(\vv_t, j_t, l_t) \mid z] =
\begin{cases}
1-\xi' & \text{if } w_z^{j_t}[l_t] = \vv_t[l_t] \\
\xi' & \text{if } w_z^{j_t}[l_t] \neq \vv_t[l_t]
\end{cases}
\]
In this paper we do not investigate the effectiveness of BPIA against HCMS, and we leave this to future work.

\subsection{Correctness of score} \label{sec:correctness_score}

We show that the estimation for $\score(P_i)$ in eq.~\ref{eq:score_estimate} is correct, i.e.\ that it is proportional to $\Pr_{\M}[P_\U = P_i \mid \x_1, \ldots, \x_n]$ under the hierarchical model $\M$. We abbreviate $\x_1, \ldots, \x_n$ with $\underline{\x}$. To clarify between random variables and their realizations, we write the first ones in uppercase, so that $\M$'s hyperparameters $\gamma$ and $\delta$ are written as $\Gamma$ and $\Delta$ respectively. Let $\Theta = (0,1] \times (1/k,1]$. Then:
\begin{align}
& \Pr_{\M}[P_\U = P_i \mid \underline{\x}] 
= \Pr_{\M}[\iota = i \mid \underline{\x}] 
\propto \Pr_{\M}[\iota = i] \Pr_{\M}[\underline{\x} \mid \iota = i] \nonumber \\
= & p_\iota(P_i) \int_\Theta \Pr_{\M}[\underline{\x} \mid \iota = i, \Gamma = \gamma, \Delta = \delta] \ d (p_\Gamma(\gamma), p_\Delta(\delta)) \label{eq_step:marginalize} \\
= & \frac{1}{k} \int_0^1 \int_{\frac{1}{k}}^1 1 \frac{1}{1-1/k} \Pr_{\M}[\underline{\x} \mid \iota = i, \Gamma = \gamma, \Delta = \delta] \ d \delta \ d \gamma \label{eq_step:uniform} \\
\propto & \int_0^1 \int_{\frac{1}{k}}^1 \prod_{t=1}^n \Pr_{\M}[\x_t \mid \iota = i, \Gamma = \gamma, \Delta = \delta] \ d \delta \ d \gamma \nonumber \\
= & \int_0^1 \int_{\frac{1}{k}}^1 \prod_{t=1}^n \sum_{z \in \Omega} \Pr_{\M}[\x_t \mid z] \ \Pr_{\M}[z \mid \iota = i, \Gamma = \gamma, \Delta = \delta] \ d \delta \ d \gamma \nonumber \\
= & \int_0^1 \int_{\frac{1}{k}}^1 \prod_{t=1}^n \sum_{z \in \Omega} \Pr_{\A}[\x_t \mid z] \ \ol{\Phi}(z \mid i, \gamma, \delta, \p_\Omega) \ d \delta \ d \gamma \label{eq_step:last_repl}
\end{align}
In eq.~\ref{eq_step:marginalize} we marginalized over the hyperparameters $\gamma$ and $\delta$, since these are not known to $\Adv$; in eq.~\ref{eq_step:uniform} we used the fact that $\iota, \gamma, \delta$ are all uniformly distributed according to $\M$ (as we are assuming a weak adversary with no knowledge on the distribution of these parameters, see also Appendix~\ref{sec:other_aux_info}); in eq.~\ref{eq_step:last_repl} we are directly using the definition of $\M$ (specifically of $X_t$ and $\X_t$).

\subsection{Improving the attack with other types of auxiliary information.} \label{sec:other_aux_info}
In the experiments with synthetic users we decide to quantify the effectiveness of the attack by randomizing the user's preferred pool and using uniform priors in the attack for $\iota$, $\gamma$ and $\delta$. In practice the adversary may use additional types of auxiliary information to estimate distributions that would likely improve the effectiveness of the attack in practice. For example, the adversary might know that the user often uses skin-toned emojis, or that they are likely to use almost always the same skin tone. This information could be incorporated into $\M$ using \emph{behavioral priors} $p_\gamma$ and $p_\delta$ for $\gamma$ and $\delta$, respectively. In another instance, the adversary might know that the target user lives in a city where the majority of the population is white, and hence might expect the target user to be more likely to use light skin-toned emojis. This information can be used to build a \emph{pool prior} $p_\iota$ for $\iota$. We leave the study of these potential improvements for future work.

\subsection{Effect of hashing}
The use of hashing in CMS increases the uncertainty on which original object (input) was used to produce the given obfuscated object (output) (see section~\ref{sec:background} and Appendix~\ref{sec:CMS}). We now show that hashing has however close to no impact on our attack. To do this, we modify CMS to remove hashing while keeping the random flipping of bits. Specifically, this means having $m = |\Omega|$ and replacing the family of hash functions $\H$ with a single bijective function $h \colon \Omega \to \{1,\ldots,|\Omega|\}$, so that there are no collisions.

Table~\ref{tab:news-AUC-PN-no_hash} shows the effectiveness of the attack in the web domains setting without hashing. Interestingly, we see that the results are almost exactly the same as with the standard CMS with hashing (Table~\ref{tab:news-AUC-PN}), meaning that hashing has a negligible effect on our attack. Intuitively, the reason is that hash collisions are distributed roughly uniformly in the universe $\Omega$. Hence, while collisions provide strong obfuscation for a single object, over multiple observations the effect of collisions tends to ``average out'' and the information on the preferred pool is leaked anyway.

We omit the results for the emojis setting as they are very similar.
\begin{table}[t]%
\begin{center}
\begin{tabular}{ccccc}
\toprule
& $n=7$ & $n=30$ & $n=90$ & $n=180$ \\
\midrule
$\Advw$     &  0.72 &  0.89 &  0.96 &  0.97 \\
$\Advs$     &  0.74 &  0.90 &  0.96 &  0.98 \\
\bottomrule
\end{tabular}
\end{center}
\caption{$\AUC$ values in the web domains setting for a modified CMS without hashing.}
\label{tab:news-AUC-PN-no_hash}
\end{table}

\end{document}